\documentclass[a4paper,12pt]{article}

\usepackage{bbold}
\usepackage{float}
\usepackage{cancel}
\usepackage{amsmath, amssymb, setspace}
\usepackage{slashed}
\usepackage{graphicx}
\usepackage[hidelinks]{hyperref}
\usepackage{cite}
\usepackage{color}
\usepackage{soul}
\usepackage{xcolor}
\usepackage[dvipsnames]{xcolor}
\usepackage{braket}
\usepackage{makecell}
\usepackage{hhline}
\usepackage[compat=1.1.0]{tikz-feynman}
\numberwithin{equation}{section}
\vspace{0.6cm}
\date{\today} 

\usepackage{tabularx}
\usepackage{longtable} 
\usepackage{multicol}
\usepackage{multirow}
\usepackage{enumitem}
\usepackage{caption}
\usepackage{subcaption}
\usepackage[normalem]{ulem}
\begin{document}

\hfill {\tt CERN-TH-2025-253, MITP-25-076, SI-HEP-2025-28,
P3H-25-109}

\begin{center}
\Large\bf\boldmath
\vspace*{1.cm} 
Nonlocal form factor of chromomagnetic penguin
in $B\to K\ell^+\ell^-$ from QCD light-cone sum rules
\unboldmath
\end{center}

\def\thefootnote{\fnsymbol{footnote}}

\vspace{0.5cm}
\begin{center}
%Name1$^{1}$, 
T.~Hurth$^{1}$, 
A.~Khodjamirian$^{2}$, 
F.~Mahmoudi$^{3,4,5}$,\\
D.~Mishra$^{3}$,
Y.~Monceaux$^{3}$, 
S. Neshatpour$^{3}$\\
\vspace{1.cm}
{\sl $^1$
PRISMA+ Cluster of Excellence and Institute for Physics (THEP),
Johannes Gutenberg University, D-55099 Mainz, Germany
}\\[0.4cm]
{\sl $^2$
Center for Particle Physics (CPPS), Theoretische Physik 1, \\Universit\"at Siegen, D-57068 Siegen, Germany
}\\[0.4cm]
{\sl $^3$Universit\'e Claude Bernard Lyon 1, CNRS/IN2P3, \\
Institut de Physique des 2 Infinis de Lyon, UMR 5822, F-69622, Villeurbanne, France}\\[0.4cm]
{\sl $^4$Theoretical Physics Department, CERN, CH-1211 Geneva 23, Switzerland}\\[0.4cm]
{\sl $^5$Institut Universitaire de France (IUF), 75005 Paris, France }\\[0.4cm]
\end{center}
\renewcommand{\thefootnote}{\arabic{footnote}}
\setcounter{footnote}{0}

\vspace{1.cm}
\begin{abstract}
The branching fraction of the $B \to K\ell^+\ell^-$ decay has been measured recently by the LHC experiments, showing a deviation from theory predictions based on the Standard Model (SM). A major challenge in achieving a complete SM prediction and interpreting this discrepancy lies in the treatment of nonlocal hadronic effects. In $B \to K\ell^+\ell^-$, these effects are cast in a single nonlocal form factor, a function of squared momentum transfer $q^2$ to the lepton pair. One of the 
previously used methods provides this form factor in the region of spacelike momentum transfer, $q^2<0$,
matching the result to the hadronic dispersion relation, which is then continued to the physical region. The calculation 
done so far was a combination of QCD factorisation for hard-gluon contributions with light-cone sum rules (LCSRs) for soft-gluon ones. In this work, we calculate for the first time the
complete nonlocal form factor at $q^2<0$ for one of the effective operators,
the chromomagnetic operator $O_{8g}$, applying the method of LCSRs 
with $ B$-meson distribution amplitudes.  We compute, both
analytically and numerically, the operator-product expansion (OPE) diagrams with
hard-gluon exchanges, analyse their structure and hierarchy, and obtain their spectral density
entering the LCSR together with soft-gluon contributions.
This study paves the way for our next task, a complete calculation of 
nonlocal $B \to K\ell^+\ell^-$ form factor
at spacelike $q^2$, including the dominant contributions of current-current operators,
known as charm-loops.

\end{abstract}

\clearpage

\tableofcontents

%color coding :   
%\thu{TH} \akh{AK}  \nm{FM}  \daya{DM} \sn{SN} \yam{YM}
\clearpage
\section{Introduction}
Exclusive flavour-changing neutral-current (FCNC) decays
of $B$-meson are facing a long-standing tension between theory and experiment. Currently, the most striking is the situation with
the simplest observable---the branching fraction of 
the $B\to K\ell^+\ell^-$ decay. The recent measurement of the CMS collaboration
\cite{CMS:2024syx} reveals a good agreement with the earlier 
LHCb data~\cite{LHCb:2014cxe} on this process. If one takes 
the most ``theoretically clean" bin $1.1~\mbox{GeV}^2<q^2<6.0$~GeV$^2$
of the momentum transfer $q$ to the lepton pair,
the average of these two measurements
lies below theory predictions based on the Standard Model (SM), with a tension in the range of $1.7\sigma$--$4.1\sigma$, depending on the choice of local form factors (see e.g., a recent 
detailed comparison presented in \cite{Hurth:2025vfx}).

In SM, the $B\to K\ell^+\ell^-$ decay 
amplitude, being relatively simple kinematically, involves a nontrivial  QCD dynamics embedded in two contributions of different origin. The dominant part of the decay amplitude is determined by the effective local operators  $O_{9,10}$ and $O_7$. The $B\to K$ hadronic matrix elements of these operators 
are reduced to local form factors that are accurately calculated:  (i) in lattice QCD at large 
$q^2$
and (ii) from  QCD light-cone sum rules (LCSRs) in the kinematically enhanced low-$q^2$ region. 
The remaining, presumably subdominant part of the 
$B\to K\ell^+\ell^-$ decay amplitude represents a complicated superposition of various long-distance nonlocal effects, expressed in the form of a single invariant function, the nonlocal form factor. Currently, for all FCNC exclusive 
$B$ decays, the main stumbling blocks limiting their precision 
analysis are these nonlocal form factors.

In the exclusive $b\to s \ell^+\ell^-$ decays, these effects  are generated by an overlap of the weak  $b\to s \bar{q} q$ transition
(where $q=u,d,s,c$) with the electromagnetic (e.m.)  
emission of a lepton pair. The dominant contribution, 
involving the $b\to s\bar{c}c$ transition, is known as a charm-loop effect.
In lattice QCD, dedicated investigations of nonlocal effects 
in $B\to K\ell^+\ell^-$ decays started only recently
(see e.g., \cite{Frezzotti:2025hif}). The existing QCD-based calculations
of these effects are mainly based on the heavy $b$-quark mass limit 
and on heavy-quark effective theory (HQET).

The first systematic approach \cite{Beneke:2001at}
to nonlocal effects in $B\to K^{(*)}\ell\ell$ was using the QCD factorisation (QCDf) in terms of the light-cone distribution amplitudes of $B$-meson (in HQET) and $K^{(*)}$ meson. The resulting factorisation formula, derived on the basis of this approach, includes, in addition to the leading-order charm-loop, also the effects of hard-gluon exchanges. Applying  QCDf in the physical region of the $B\to K^{(*)} \ell^+\ell^-$ decay, 
e.g. for the bin $1.1<q^2<6.0$ GeV$^2$, one has to assume, 
first of all, that a quark-level calculation of the charm-loop is directly
applicable in the timelike region.
Also, by now QCDf cannot be extended to subleading powers in
the heavy mass scale. In particular,
the soft-gluon effects generating such 
power corrections to the factorisation formulas are not yet accessible.

In \cite{Khodjamirian:2010vf}, a different approach was developed, shifting the calculation of nonlocal form factors into the region of spacelike values of $q^2$, far below hadronic thresholds.
This enables to safely compute the contributions of soft gluons emitted by the charm-loop, applying LCSRs with $B$-meson distribution amplitudes (DAs).
The other new element of this approach is the extensive use of the hadronic dispersion relation in the variable $q^2$ for a nonlocal form factor. 
Matching the parameters of this relation to  the calculated form factor  at spacelike $q^2$, one then continues the latter in the form of a dispersion relation into the physical region of $q^2$, avoiding a direct use of quark-level
expressions in the timelike region. The price for that is an inevitable dependence of the result on the chosen ansatz for the resonances structure of the dispersion relation. 

Following this approach, a complete calculation of nonlocal form factor for $B\to K \ell^+\ell^-$ was accomplished in 
\cite{Khodjamirian:2012rm} (see also 
\cite{Khodjamirian:2017fxg}). The contributions of all effective operators and all quark topologies were included, combining
QCDf for the perturbative contributions 
with LCSRs for the $B\to K$ form factors and for the soft-gluon contributions. In \cite{Bobeth:2017vxj,Gubernari:2020eft,
Gubernari:2022hxn}, this method was extended to the $B\to K^*\ell^+\ell^-$ mode and further developed, adding dispersive bounds.

Facing a noticeable discrepancy between the predicted and measured $B\to K\ell^+\ell^-$ branching fractions, it is timely to carefully revise the methods used to evaluate nonlocal effects. In particular, it is desirable to reanalyse the approach mentioned above that combines LCSR and QCDf at spacelike $q^2$ with hadronic dispersion relation in the physical region.
Before turning to the phenomenological component of that approach, which is 
mainly the dispersion relation, it is necessary to upgrade the QCD-based component.
To this end, we plan to undertake a systematic computation 
of all contributions to the  $B\to K\ell^+\ell^-$ nonlocal form factor at spacelike $q^2$ within one and the same LCSR method and with a uniform nonperturbative input in a form of $B$-meson DAs. 
This program is, however, technically very demanding, mainly because
the operator-product expansion (OPE)  
of the underlying correlation function 
used for LCSR contains two-loop diagrams with several mass scales.

The aim of this paper is to accomplish the first step in this direction
and to prove that our program is feasible  for a selected contribution to 
the $B\to K\ell^+\ell^-$  nonlocal form factor.
To this end, we choose the chromomagnetic operator $O_{8g}$ which does not contain any additional energy/mass scale, such as $c$-quark mass inherent for the charm-loop. With this choice, we avoid too complicated two-loop diagrams related to the hard-gluon exchanges between the charm-loop and other quarks 
in that process. On the other hand, calculating the nonlocal form factor for $O_{8g}$ is a convenient starting point because we will reveal the relative importance of separate diagrams depending on their topology and on the location of the virtual photon emission point. We will also reproduce properties of the nonlocal form factor, such as the imaginary part emerging already at $q^2<0$. 

The paper is organised as follows. In Section~\ref{sec:O8g_NLff}, we introduce the nonlocal form factor for $B\to K\ell^+\ell^-$ associated with the chromomagnetic operator $O_{8g}$. Section~\ref{sec:lcsr} presents the derivation of the corresponding LCSRs for this nonlocal form factor. In Section~\ref{sect:opediags}, we describe the computation of the OPE diagrams, beginning with the nomenclature of diagrams and proceeding through the hard-factorisable, hard nonfactorisable, and soft nonfactorisable topologies, followed by the discussion on power counting. Section~\ref{sec:results} contains the numerical analysis of our results, and Section~\ref{sec:discussions} provides a summary of the main findings and an outline of the future plans.

\section{Nonlocal form factor of the operator \texorpdfstring{$O_{8g}$}{O8g}}
\label{sec:O8g_NLff}
In the SM, the effective 
Hamiltonian of the $b\to s$  transitions 
contains, among other operators, also 
the chromomagnetic operator $O_{8g}$ with its Wilson coefficient $C_{8g}$. This local,
dimension-six operator emerges after integrating out virtual $t$-quarks and $W$-bosons in the loop diagrams where a single gluon is emitted.  We use the standard form:
\begin{eqnarray}
    O_{8g}=-\frac{g_s}{16\pi^2}\,m_b\,\bar{s}\,\sigma^{\mu\nu}P_R\,G_{\mu\nu}\,b\,,
\label{eq:C8g}
\end{eqnarray}
where $P_R\equiv (1+ \gamma_5)/2$ and the suppressed part proportional to $m_s$ is neglected (the convention used for $\sigma^{\mu\nu}$ is given in Appendix~\ref{app:notation}). 

We aim at calculating the matrix element in which a time-ordered product 
of the operator $O_{8g}$ with the quark e.m. current 
\begin{equation}
j_\mu^{em}=\sum_{q=u,d,s,c,b}Q_q\bar{q}\gamma_\mu q\,,\label{eq:jem}
\end{equation}
is sandwiched between the
$B$-meson and kaon on-shell states:
\begin{eqnarray}
{\cal H}^{(BK,O_{8g})}_{\mu}(p,q)= i\int d^4x\, e^{iq\cdot x}\, \langle \bar{K} (p)|T\{j_\mu^{\rm em}(x),\,C_{8g}O_{8g}(0)\}|\bar{B}(p_B)\rangle
\nonumber \\
=\left( (p \cdot q)\, q^\mu - q^2 p_\mu \right) 
\mathcal{H}^{(BK,\,O_8)}(q^2)
\,,
\label{eq:hme}
\end{eqnarray}
where the four-momentum transfer is $q=p_B-p$ and the  $B$-meson and kaon momenta are on-shell, $p_B^2=m_B^2$, $p^2=m_K^2$ .
In the r.h.s. of (\ref{eq:hme}), we use the e.m. current conservation, allowing 
us to reduce this hadronic matrix element to a single invariant function of the momentum transfer squared, that is the nonlocal form factor denoted as ${\cal H}^{(BK,\,O_8)}(q^2)$.

Since we are not considering the full $B\to K \ell^+\ell^-$  decay amplitude in this paper, and, in particular, we ignore the ``direct" operators $O_9, O_{10}$ and $O_7$, it is more convenient to treat the matrix element (\ref{eq:hme}) as a part of the $B\to K\gamma^*$ amplitude with a virtual photon emission. This interpretation also makes it easier
to consider the spacelike region of momentum transfer, $q^2<0$ which is 
unphysical for the decay into a lepton pair,
but plays a major role here, being accessible 
with our calculational method.

The amplitude (\ref{eq:hme}) with the operator $O_{8g}$ plays a minor role compared
to the dominant nonlocal effects in $B\to K\gamma^*$ transitions, 
stemming from the intermediate $c$-quark loop.  
These ``charm-loop" effects emerge 
due to the $b\to c \bar c s $ operators $O_{1,2}^{(c)}$  in the effective Hamiltonian with their Wilson coefficients  $C_{1,2}$ larger than
$C_{8g}$.
Still, the coefficient of the chromomagnetic operator itself is much larger than Wilson coefficients of the quark-penguin or electroweak-penguin operators, which we neglect hereafter.\footnote{ Hence, we also neglect the mixing effects between operators for simplicity.}

To quantify the $O_{8g}$ effect in the total budget of nonlocal contributions to $B\to K\gamma^*$, we introduce the ratio
of this effect to the leading-order charm-loop with a virtual photon emission, taken in the factorisable approximation. Using the expression for the latter
from~\cite{Khodjamirian:2012rm}, we define
\begin{equation}
R_{(O_{8}/O_{1,2}^{c})}(q^2)\equiv \frac{\mathcal{H}^{(BK,\,O_8)}(q^2)}{
\mathcal{H}^{(BK,\,O_{1,2}^c)}(q^2)}\,,
\label{eq:ratioO8}
\end{equation}
where 
\begin{eqnarray}
{\mathcal H}^{(BK,\,O_{1,2}^c)}(q^2) = \frac{Q_c}{8 \pi^2}
\left(\frac{C_1}3+C_2\right)g(m_c^2,q^2) f^+_{BK}(q^2)\,,
\label{eq:Afact}
\end{eqnarray}
and
\begin{eqnarray}
g(m_q^2,q^2) &=& -
\left (\ln {m_q^2 \over \mu^2} +1
\right ) + q^2    \int\limits_{4 m_q^2}^{+\infty} ds   {
\sqrt{1- {4 m_q^2 \over s}} \left ( 1+ { 2 m_q^2 \over s} \right)
\over  s (s-q^2)} \,,
\label{eq:gfunct}
\end{eqnarray}
is the well-known loop function, $\mu$ is the renormalisation scale
and $f^+_{BK}(q^2)$ is the $B\to K$ local form factor of the vector $b\to s$ 
current. 

\section{Light-cone sum rule for the nonlocal form factor}
\label{sec:lcsr}

We use the method of LCSRs with $B$-meson DAs in HQET, 
that was initially developed in~\cite{Khodjamirian:2006st} (also in the framework of soft-collinear effective theory (SCET) in \cite{DeFazio:2005dx})\,, 
to obtain the $B\to$ light meson form factors,
including also the $B\to K$ local form factors 
(see \cite{Khodjamirian:2023wol}  for a detailed review of
more recent results).
Following \cite{Khodjamirian:2012rm}, we start from the correlation function:
\begin{eqnarray}
    \Pi_{\nu\mu}(p,q) &=& i^2\int d^4y\, e^{ip\cdot y}\,\int d^4x\, e^{iq\cdot x}\, \langle 0|T\{
    j_{\nu 5}^{K}(y)\,
    j_\mu^{\rm em}(x)\, C_{8g}\, O_{8g}(0)\}|\bar{B}(p_B)\rangle\,
    \nonumber\\
&=&i\, p_\nu \, q_\mu \Pi(p^2,q^2)\, + \, \dots\, ,
\label{eq:corr}
\end{eqnarray}
where we select a single 
kinematical structure from the Lorentz-expansion, 
so that the  invariant amplitude $\Pi(p^2,q^2)$
will be used in the resulting LCSR. The other structures are irrelevant and denoted by ellipsis.
In the correlation function of our choice 
the kaon is interpolated by the axial current, 
\begin{eqnarray}
    j_{5\nu}^{K}(x)=\bar{q}(x)\gamma_\nu\gamma_5\,s(x)\,,
\label{eq:jkaon}
\end{eqnarray}
with $q = u,d$ for the charged and neutral kaon, respectively.

Considering the correlation function (\ref{eq:corr}) as an analytic function of the variable $p^2$ at fixed spacelike $q^2$, we derive a dispersion relation in the kaon channel. To this end, we apply unitarity condition, 
inserting in  (\ref{eq:corr}) a complete set of intermediate hadronic states carrying kaon quantum numbers. Isolating the kaon pole, we have:
\begin{equation}
\Pi_{\nu\mu}(p,q)= \frac{\langle 0|j_{5\nu}^{K}|\bar{K}(p)\rangle \,
\,
i\int d^4x\, e^{iq\cdot x}\,
\langle \bar{K}(p)|
T\{
j_\mu^{\rm em}(x)\, C_{8g} O_{8g}(0)\}|\bar{B}(p_B)\rangle}{m_K^2-p^2} 
+\int\limits_{s_h}^\infty ds\,\frac{\rho_{\nu\mu}^{\rm h}(s,p,q)}{s-p^2}\,,
\label{eq:had_disp}
\end{equation}
where the spectral density 
$\rho_{\nu\mu}^{\rm h}(s,p,q)=ip_\nu q_\mu\rho^{\rm h}(s,q^2)+ ...$,
in which we isolate the relevant invariant function,
is formed by the sum over contributions of all heavier
than kaon hadronic states with the quantum numbers
of the $j_{5\nu}^K$ current. The lightest in this sum is the $K\pi\pi$ continuum state, hence  $s_h=(m_K+2m_\pi)^2$.
Applying the standard definition for the decay constant of the kaon: 
\begin{equation}
\langle 0|j_{5\nu}^{K}|\bar{K}(p)\rangle = i\,f_K\,p_\nu\,,
\label{eq:fK}
\end{equation}
and the definition  (\ref{eq:hme}) of the  nonlocal form factor,
we obtain from (\ref{eq:had_disp}) the dispersion relation for the invariant amplitude: 
\begin{align}
\Pi(p^2,q^2)
= \frac{f_K (m_B^2-q^2-m_K^2)\mathcal{H}^{(BK,\,O_8)}(q^2)}{2(m_K^2-p^2)} +\int\limits_{s_h}^\infty ds\,\frac{
\rho^{\rm h }(s,q^2)}{s-p^2}\,,
\label{eq:hadrdisp}
\end{align}
where we replace $p^2\to m_K^2$ in  the kinematical factor 
$(p\cdot q)$ to remove
the constant term that will vanish after Borel transform.

In the following section, we compute the 
correlation function (\ref{eq:corr}) applying the light-cone OPE
in terms of $B$-meson distribution amplitudes.
In the OPE result, we isolate the relevant invariant amplitude: 
 $$\Pi^{\rm (OPE)}_{\nu\mu}(p,q)=i p_\nu q_\mu \Pi^{\rm (OPE)}(p^2,q^2)+ \dots\,,$$
transforming it to the form of a dispersion integral in the variable~$p^2$:
\begin{equation}
 \Pi^{\rm (OPE)}(p^2,q^2) =  \frac{1}{\pi}\int\limits_{m_s^2}^\infty ds\,\frac{{\rm Im}\Pi^{\rm (OPE)}(s, q^2)}{s-p^2}\,.
 \label{eq:OPEdispform}
\end{equation}
In the spacelike $p^2$-region of the OPE validity, we equate
the dispersion relation (\ref{eq:hadrdisp})
to the amplitude $\Pi^{\rm (OPE)}(p^2,q^2)$.
The dispersion form of the latter amplitude allows us to apply the usual quark-hadron duality approximation for the integral over the heavier state
in the dispersion relation (\ref{eq:hadrdisp}):
\begin{equation}
\int\limits_{s_h}^\infty ds\,\frac{
\rho^{\rm h }(s,q^2)}{s-p^2}=
\frac{1}{\pi}\int\limits_{s_0}^\infty ds\,\frac{{\rm Im}\,\Pi^{\rm (OPE)}(s, q^2)}{s-p^2}\,,
\label{eq:dual}    
\end{equation}
where $s_0$ is the effective threshold. 
After that, we substitute (\ref{eq:OPEdispform}) to the l.h.s. of (\ref{eq:hadrdisp}) and replace the integral over $\rho^{\rm h}(s)$ 
by the duality approximation (\ref{eq:dual}). After subtracting
equal integrals from both parts of the resulting equation 
and after a standard Borel transform, we obtain the desired LCSR
for the nonlocal form factor:
\begin{equation}
{\mathcal H}^{(BK,\,O_8)}(q^2) 
= \frac{2\, e^{m_K^2/M^2}}{f_K\, (m_B^2 - m_K^2 -q^2)} \frac{1}{\pi} 
\int\limits_{m_s^2}^{s_0} ds\,e^{-s/M^2}\, {\rm Im}\,\Pi^{\rm (OPE)}(s,q^2)\,.
\label{eq:lcsr}
\end{equation}
This sum rule was earlier derived also in \cite{Khodjamirian:2012rm}
but the OPE calculated there was limited to a single specific contribution
due to the soft-gluon absorbed in three-particle DAs of $B$-meson.
The most important parts with hard-gluon were estimated using QCDf results.
In what follows, we will accomplish, for the first time the complete computation of OPE including both hard-gluon and soft-gluon parts within one and the same method.

\section{Computation of OPE diagrams}  
\label{sect:opediags}
\subsection{Nomenclature of diagrams}

\begin{figure}[h]
	 	\begin{subfigure}{.5\textwidth}
	 		\centering
	 		\includegraphics[width=0.9\textwidth]{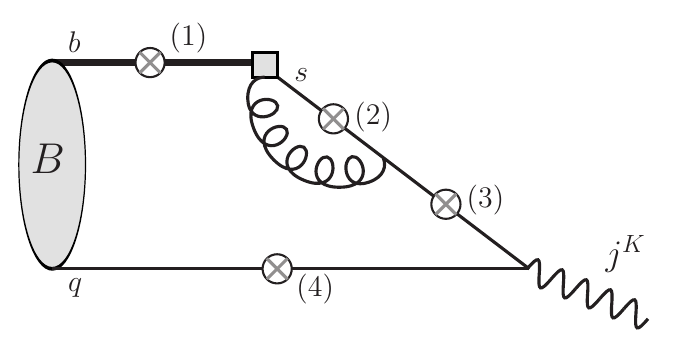}
	 		\caption{}
            \label{fig:fact_HG_A}
	 	\end{subfigure}% 	
         \begin{subfigure}{.5\textwidth}
	 		\centering
	 		\includegraphics[width=0.9\textwidth]{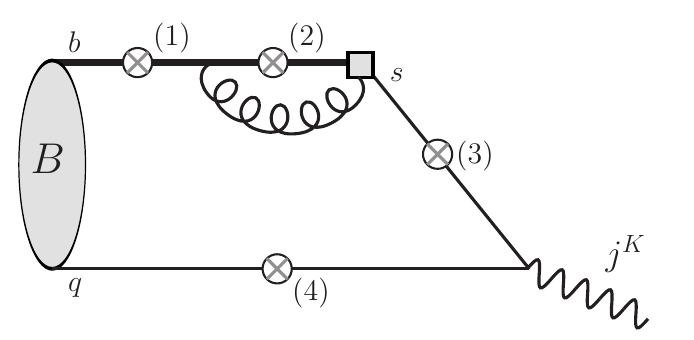}
	 		\caption{}
            \label{fig:fact_HG_B}
	 	\end{subfigure}\\
        \begin{subfigure}{.5\textwidth}
	 		\centering
	 \includegraphics[width=0.9\textwidth]{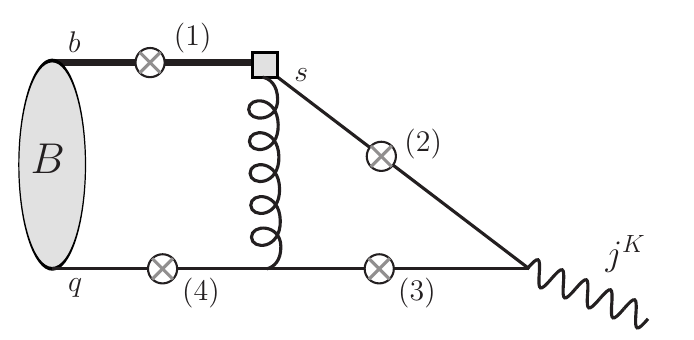}
	 		\caption{}
            \label{fig:nonfact_HG}
	 	\end{subfigure}% 	
        \begin{subfigure}{.5\textwidth}
	 		\centering
	 \includegraphics[width=0.9\textwidth]{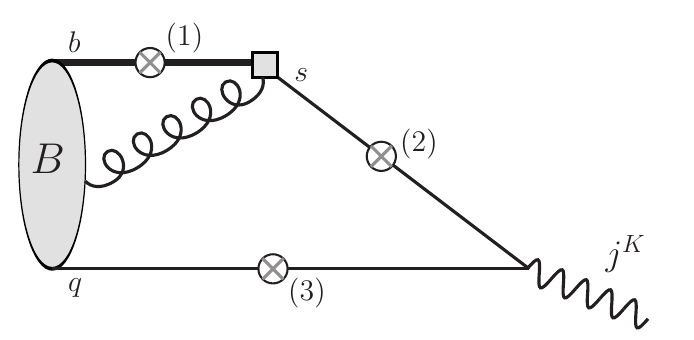}
	 		\caption{}
            \label{fig:nonfact_SG}
	 	\end{subfigure}
	 	\caption{
        Diagrams for the correlation function (\ref{eq:corr}). The 
        chromomagnetic operator $O_{8g}$ is 
        denoted by a shaded square, the 
        $B$-meson DAs are shown with a shaded blob.
        The circled crosses indicate  possible emission points of a virtual photon.  }
	 	\label{fig:O8gdiag}
	 \end{figure}

In Fig.~\ref{fig:O8gdiag}, the diagrams corresponding to the OPE of the correlation function (\ref{eq:corr}) are depicted.
For perturbative gluons, we confine ourselves to the leading-order, $O(\alpha_s)$ accuracy, so that they are emitted from the operator
$O_{8g}$ and absorbed by one of the three quark lines.
Altogether, there are three sets of one-loop diagrams denoted by $(A),(B)$ and $(C)$. Each set is characterised by a definite quark-gluon topology, and contains four separate
diagrams corresponding to four possible ways to insert the e.m. current, or, in other words, to emit a virtual photon from quark lines. For the 
sake of compactness,
we show in a single subfigure all four diagrams of the same topology, numbering them according to the e.m. current insertion point, counted clockwise. 
 For instance, $(A1)$ is the diagram of type $A$
where the e.m. current is inserted in the 
virtual $b$-quark line. 

The diagrams $(A),(B),(C)$ with hard gluon exchanges are described by convolutions of the short-distance perturbative kernels with the
two-particle, quark-antiquark $B$-meson DAs. In these DAs, we take into account the leading and subleading twist-2 and twist-3 components. Here we follow the twist expansion of the $B$-meson DAs worked out in  \cite{Braun:2017liq}.

The gluon emitted from the $O_{8g}$ operator vertex 
in the correlation function can also have a low virtuality of $O(\Lambda_{QCD}^2)$. Such a ``soft" gluon,
together with the heavy-light quark-antiquark pair, forms
 $B$-meson three-particle
DAs. There are  three soft-gluon diagrams  as indicated in 
the subfigure $(D)$ of Fig.~\ref{fig:O8gdiag}. They correspond to the three possible ways to insert the e.m. current. In the OPE, the three-particle DAs start from twist-3. That will be our default choice,
consistent with the adopted twist-3 accuracy of  
the two-particle DAs. To investigate the sensitivity to higher twists, we will also include 
into the soft-gluon part of the correlation function
all three-particle DAs up to twist-6. Note that in these 
diagrams, both the gluon and quark entering the $B$-meson DA are emitted from a single point, making the light-cone expansion with respect to this emission automatically local, and 
leaving only one interval $x^2$ which has to be close to the light-cone.
One of the advantages of the LCSR method is that the complete correlation function 
can be separated into a short-distance perturbative part and a
long-distance part represented by $B$-meson DAs.
This factorisation property is valid for  each diagram in Fig.~\ref{fig:O8gdiag}
including hard-gluon and soft-gluon contributions.
 In this paper we do not go further into more detail in that direction, 
and will not dwell on the renormalisation effects and separation-scale dependence.

In the QCDf approach \cite{Beneke:2001at} to 
$B\to K \ell^+\ell^-$ decay, if we only retain the $O_{8g}$ operator, 
there are two distinct types of
diagrams. One is represented by perturbative gluon exchanges adjacent to the operator vertex, so that the long-distance part can be reduced to a local $B\to K$ form factor. The other type of diagrams contains hard-scattering, that is, the exchange of perturbative gluon between the operator vertex and spectator quark in $B$-meson. We note that  some of the diagrams in Fig.~\ref{fig:O8gdiag} have their counterpart in QCDf, if we, conditionally, replace the kaon interpolating current by the kaon DA. In particular, diagrams $(A1)$--$(A3)$, $(B1)$--$(B3)$ can be  reduced to the local form factors, whereas diagrams $(C1)$--$(C4)$ can be classified as hard-scattering ones.
However, diagrams ($A4$) and ($B4$), where the virtual photon is emitted from the spectator quark, with no gluon exchange between quarks,  have no analogs among the QCDf contributions. In fact, we shall see that in our LCSR these two diagrams, as well as diagram $(D3)$ do not contribute, for the reason to be explained in detail below.  
Furthermore, diagrams $(D1)$--$(D3)$ with soft gluons are also absent in the present 
calculations based on QCDf, because in that method a factorisation formula at subleading power is not established yet. 

Accordingly, in what follows, we will
separate from each other, the form factor-like,
hard-scattering and soft contributions to the LCSR 
(\ref{eq:lcsr}), stemming, respectively, from the diagrams $\{(A1)$--$(A3),(B1)$--$(B3)\}$,  $\{(C1)$--$(C4)\}$ and  $\{(D1)$--$(D2)\}$.
In addition, in each of these three parts we will also collect  contributions in which 
a virtual photon is emitted from a
quark line with a certain flavour, in our case $b$,\,$s$ or $q=u,d$. These flavour-specific contributions will be 
distinguished by their global quark-charge  factor $Q_b$, $Q_s$ or $Q_{u,d}$, respectively,

The task to evaluate the OPE diagrams is somewhat simplified by the fact that only the imaginary part of the invariant amplitude $\Pi(p^2,q^2)$
in the variable $p^2$ contributes to the LCSR (\ref{eq:lcsr}).
A direct computation, to be discussed in detail  in the following subsections, reveals  that the diagrams $(A4), (B4)$ and $(D3)$ develop imaginary part  at large $p^2=s$,
far above a typical duality interval $s< s_0\sim O(1\, \mbox{GeV}^2)$ for the kaon, hence, these three diagrams do not contribute to the LCSR. 

The shift of imaginary part has in fact a qualitative
explanation based on kinematics\footnote{ A similar consideration for a different QCD
sum rule can also be found in \cite{Bordone:2022drp}.}. 
Taking the imaginary part of the diagrams $(A4),(B4),(D3)$ 
in the variable $p^2$, we put on-shell the intermediate light-quark line between the
kaon and e.m. current vertices. The $s$-quark line in these diagrams remains 
off-shell, because it absorbs the initial momentum $p_b$ of the quasi-on-shell $b$-quark
line in the $B$-meson (or the sum of $b$-quark and soft gluon momenta in the diagram
$(D3)$). The fact that there is an intermediate quark-gluon loop in the diagrams $(A4),(B4)$ does not play a role, if only kinematics is concerned. In the rest frame of the $B$-meson, $p_B=(m_B,\vec{0})$, we have two pairs
of back-to-back three-momenta: first,  $\vec{p}_b=-\vec{p}_q$  for the $B$-meson constituents and second, $\vec{p}=-\vec{q}$ for the kaon-interpolating and e.m. currents. Energy-momentum conservation tells that in this frame:
\begin{equation}
|\vec{p}_b|=|\vec{p}_q|=\frac{m_B^2-m_b^2}{2m_B}\simeq \bar\Lambda, ~~~~  
|\vec{p}\,|=|\vec{q}\,|=\frac{\lambda^{1/2}(m_B^2,q^2,p^2)}{2m_B}\,,
\label{eq:vect}
\end{equation}
where in the first equation the HQET parameter $\bar{\Lambda}=m_B-m_b$ is introduced
and terms of $O(\bar{\Lambda}^2)$ are neglected, and in the second equation $\lambda$ is the K\"{a}ll\'{e}n function. At the same frame,
the four-momentum $f$ of the light-quark line
between currents has the following components:
\begin{equation} 
f_0=\frac{m_b^2-p^2+q^2}{2m_B}, ~~ ~
|\vec{f}\,|=|\vec{p_b}\,|^2-2|\vec{p}_b\,||\vec{p}\,|\cos\theta +|\vec{p}\,|^2\,,
\label{eq:fcomp}
\end{equation}
where $\theta$ is the angle between $\vec{p}_b$ and $\vec{p}$. Furthermore, due to 
(\ref{eq:vect}), these components are expressed via
invariant variables $p^2$ and $q^2$.
Hence, the on-shell condition 
$f^2=f_0^2-\vec{f}^{\,2}=0$  yields a relation between 
these variables: 
\begin{equation} 
q^2+\frac{\overline{\Lambda}}{m_B}\big[(m_B^2-p^2+q^2)^2-4m_B^2q^2\big]^{1/2} 
\cos \theta -O(\overline{\Lambda}^2)=0 \,.
\label{eq:p2cond}
\end{equation}
This relation
can only be satisfied if the variable $p^2 \gtrsim  \mathcal{O}(m_B^2+(-) q^2)$ at 
nonzero $\cos\theta>(<)0$. Note that in the region of spacelike $q^2 < 0$, 
where LCSRs will be used, we have, parametrically,
$\Lambda_{QCD}^2\ll |q^2|\ll m_b^2$. Hence, the corresponding values of $p^2$ are much larger than $m_K^2$, implying that the pole in the $p^2=s>0$ region for the diagrams
$(A4),(B4),(D3)$ is located far above a typical duality interval for the kaon current.

Taking into account the separation of contributions
into form-factor-like $(ff)$, hard scattering $(hs)$ and 
soft-gluon $(sg)$
as suggested above, the OPE result to be put on r.h.s. of LCSR (\ref{eq:lcsr}) takes the following form 
\begin{align}
   \frac{1}{\pi} \int\limits_{m_s^2}^{s_0} ds\,e^{-s/M^2}\, {\rm Im}\, \Pi^{\rm (OPE)}(s,q^2) &= {\cal P}^{\rm (OPE)}_{(ff)}(M^2,q^2) + {\cal P}^{\rm (OPE)}_{ (hs)}(M^2,q^2) + {\cal P}^{\rm (OPE)}_{(sg)}(M^2,q^2)\, , 
\end{align}
where the three terms above are further split into
separate diagram contributions, factoring  the quark charges 
for convenience:
\begin{eqnarray}
    {\cal P}^{\rm (OPE)}_{(ff)} (M^2,q^2) \!\! &=&\!\!
f_B m_B\!\left(\frac{\alpha_s C_F}{32\pi}\right)\!  \,m_b\, C_{8g}\Bigg\{
    Q_b \bigg[ I^{(A1)}(M^2,q^2) + I^{(B1)}(M^2, q^2) + I^{(B2)}(M^2,q^2)\bigg] \nonumber\\ 
\!\!\!&+&\!\!\!Q_s\ \bigg[  I^{(A2)}(M^2,q^2) + I^{(A3)}(M^2,q^2) +  I^{(B3)}(M^2,q^2)  \bigg] 
    \Bigg\} \,,
    \label{eq:Phf}
\\
{\cal P}^{\rm (OPE)}_{(hs)} (M^2,q^2)\!\! &=&\!\!  
f_B m_B \!\left(\frac{\alpha_s C_F}{32\pi}\right) \! \,m_b \, C_{8g}\,    
       \Bigg\{ Q_b I^{(C1)}(M^2,q^2) + Q_s I^{(C2)}( M^2,q^2) 
       \nonumber\\
\!\!\!&+&\!\!\! Q_d\,\bigg[I^{(C3)}(M^2, q^2) + I^{(C4)}(M^2,q^2) \bigg] \Bigg\}\,, 
\label{eq:Phnf}
\\
{\cal P}^{\rm (OPE)}_{(sg)}(M^2,q^2)\!\! &= &\!\!
f_B\frac{m_b}{8\pi^2} \,C_{8g} \,
    \Bigg\{Q_b\ I^{(D1)}(M^2,q^2) + Q_s\ I^{(D2)}(M^2,q^2)\Bigg\}\,.
\label{eq:Psnf}
\end{eqnarray}
In the following subsections we will discuss the computation of separate diagrams, resulting in the 
analytic expressions for the integrals $I^{(A1),...,(D2)}$ entering 
(\ref{eq:Phf}), (\ref{eq:Phnf}) and (\ref{eq:Psnf}).

\subsection{Form-factor-like diagrams}
\label{sec:factodigs}

Here, we discuss in detail how diagrams $(A1)$--$(A3)$ and $(B1)$--$(B3)$ in Fig.~\ref{fig:O8gdiag} are computed.
As explained above, their contributions are conditionally identified as ``form-factor-like" ones. 
In these diagrams, the perturbative gluon from the $O_{8g}$ operator forms a loop subgraph, while the $B$-meson spectator quark  remains intact. The soft physics is then entirely encoded in the two-particle $B$-meson light-cone distribution amplitudes (LCDAs).  

As an illustrative example, we describe
in more detail the computation of diagram~($A1$), where the virtual photon is emitted from the  $b$-quark line. 
The corresponding initial expression,
with all quark and gluon fields  in the diagram shown explicitly, reads:
\begin{align}
    \Pi_{\nu\mu}^{(A1)} (q,p) &= -i^4 \int d^4y\ e^{i p\cdot y} \int d^4x\ e^{i q\cdot x} \int d^4 z \bra{0} T\Big\{[\bar{d} \gamma_\nu \gamma_5 s](y), g_s A^\sigma(z) [\bar{s} \gamma_\sigma T^a s](z),\nonumber\\&
    \frac{g_s m_b}{16\pi^2} C_{8g} k^\eta A^\rho(0) T^a [\bar{s} \sigma_{\rho\eta} (1+\gamma_5) b](0), Q_b [\bar{b} \gamma_\mu b](x) \Big\} \ket{B(p_B)}\,,
  \end{align}
where, as before, $p_B=p+q$ 
and $p_B^2=m_B^2$. Contracting the quark, antiquark and gluon fields into the propagators in  
accordance with the topology of diagram $(A1)$, we replace the emerging $B\to$ vacuum matrix element of a bilocal heavy-light quark-antiquark operator by the two-particle $B$-meson DAs, using the definition (\ref{eq:2pLCDAs}). 

The result, integrated over the coordinates then reads:
\begin{align}
\Pi_{\nu\mu}^{(A1)} (q,p)
&= -i\, f_B m_B \left(\frac{\alpha_s C_F}{32\pi}\right)m_b\,C_{8g}\, Q_b  
   \int\limits_0^{\infty} d\omega \int \frac{d^4 k}{(2\pi)^4} \nonumber \\[-6pt]
&\quad \times \Bigg\{ \frac{{\rm Tr}\!\big[ \hat{\mathcal{A}}_{\nu\eta\mu} \, \gamma_5 \big] k^\eta}{\mathcal{D}_{\!A1}} \, \phi_B^+(\omega)
   + \frac{ {\rm Tr}\!\big[ \hat{\mathcal{A}}_{\nu\eta\mu} \, \gamma^\alpha \gamma_5 \big] k^\eta}{2 \mathcal{D}_{\!A1}} \,
     \overset{\leftarrow}{\partial}_\alpha \Phi_B^\pm(\omega) \Bigg\}
     \nonumber \\[-6pt]     
     &\quad = 
     ip_\nu q_\mu\Pi^{(A1)}(p^2,q^2)+ ...\,,
  \label{eq:intA1}
\end{align}
where we use short-hand notations for the Dirac structure:
\begin{align*}
\hat{\mathcal{A}}_{\nu\eta\mu} &\equiv  \gamma_\nu \gamma_5 (\omega \slashed{v} - \slashed{p}-m_s) \gamma_\rho 
                (\omega\slashed{v}- \slashed{p} +\slashed{k}-m_s) \sigma_{\rho\eta} (1+\gamma_5) 
                \left( m_b \slashed{v}  - \slashed{q} + m_b\right) \gamma_\mu (1 + \slashed{v})\,,
\end{align*}
and for the denominator:
\begin{equation*}
    \mathcal{D}_{\!A1} \equiv (\omega v - p)^2 \, \big((m_b v  - q)^2 - m_b^2\big) \, k^2 \, (\omega v - p + k)^2\,,
\end{equation*}
where we neglect $m_s^2$ in the $s$-quark propagators. In the above, $C_F = (N_c^2 - 1)/(2N_c)$ is the standard colour factor, and $v=p_B/m_B$ is the velocity four-vector. To keep compact the second term in the above, we introduce the four-vector $\ell^\alpha \equiv \omega\, v^\alpha$ and define the differential operator  
$\overset{\leftarrow}{\partial}_\alpha\equiv\overset{\leftarrow}{\partial}/\partial \ell^\alpha$
acting on the hard-scattering kernel. 

The loop momentum integration in (\ref{eq:intA1}) is from this point onward performed in $D = 4 - 2\varepsilon$ dimensions. Both ultraviolet (UV) and infrared (IR) divergences arising from the loop integrals are dimensionally regularised. Note that the $1/\varepsilon$ terms appear only in polynomial structures in $p^2$, which are removed upon Borel transform in this variable. Note also that we only retain linear terms in the $s$-quark mass $m_s$, while the ${\cal O}(m_s^2)$ and higher-order terms are neglected, being numerically insignificant.

The expression in (\ref{eq:intA1}) has a typical factorised structure: the long-distance part represented by the $B$-meson distribution amplitudes $\phi_B^+(\omega)$ and $\Phi_B^\pm(\omega)$ is 
convoluted with the hard kernel represented by the coefficients containing traces and denominator factors. After performing the loop momentum integration, the invariant amplitude reduces to: 
\begin{align}
    \Pi^{(A1)} (p^2, q^2) =&-f_B \frac{m_B}{m_b}\!\left(\frac{\alpha_s C_F}{32\pi}\right)\, C_{8g}Q_b   \int_0^{\infty} d\omega   \Bigg\{ \frac{1}{(\omega v - p)^2} \int_0^1 dx \Bigg[  \big( \tilde{B}_0^{(0,\phi)} \mathcal{I}_{2, 0}\ \nonumber \\[-8pt] 
    &+\tilde{B}_1^{(0,\phi)} \mathcal{I}_{2, 1}) \phi_B^+(\omega)    + \Bigg( \Big( \tilde{B}_0^{(1,\Phi)}  + \tilde{B}_0^{(2,\Phi)}\Big)  \mathcal{I}_{2, 0} + \Big( \tilde{B}_1^{(1,\Phi)}  + \tilde{B}_1^{(2,\Phi)}\Big)  \mathcal{I}_{2, 1}  \nonumber \\[-8pt]
    & + \frac{\tilde{B}_0^{(3,\Phi)}}{(\omega v- p)^2} \mathcal{I}_{2, 0}   + \frac{\tilde{B}_1^{(3,\Phi)}}{(\omega v- p)^2} \mathcal{I}_{2, 1} +  x(\tilde{B}_0^{(4,\Phi)} \mathcal{I}_{3, 0} + \tilde{B}_1^{(4,\Phi)} \mathcal{I}_{3, 1}) \Bigg)\frac{\Phi_B^\pm (\omega)}{2} \Bigg] \Bigg\} \,,
\label{eq:PiA1res}
\end{align}
where the coefficients 
$$\tilde{B}_i^{(j,\phi(\Phi))}=
\tilde{B}_i^{(j,\phi(\Phi))} (p^2, q^2, m_B, m_b, m_s, x, \omega), ~~~(i=0,1;j=0,..,4) \,, $$ 
are polynomial functions of the external momenta squared, of the quark and $B$-meson masses, and of the Feynman parameter $x$. 
Their expressions are given in Appendix~\ref{app:invfuncs}. 
The loop-momentum integrals multiplied by these coefficients are defined as: 
\begin{align}
    \mathcal{I}_{n, r} &= \int \frac{d^D k}{i(2\pi)^D } \frac{(k^2)^r}{(k^2 - \Delta_{A1})^n}\,,
\end{align}
using the following notation: 
\begin{align}
\Delta_{A1} & \equiv  x(x - 1)(\omega v - p)^2 = x(x-1) (1 - \hat{\omega})(p^2 - \tilde{s})\,,
\label{eq:Delta}
\end{align}
 where
$$\hat{\omega} \equiv \frac{\omega}{m_B}, ~~~~
\tilde{s} = \hat{\omega} m_B^2 - \frac{\hat{\omega}}{1 - \hat{\omega}} q^2\,.$$
In what follows, we also use notation for the duality threshold 
in terms of the variable $\hat{\omega}_0$:
$$ \hat{\omega}_0= \frac{s_0 +m_B^2 -q^2 - \sqrt{(s_0 + m_B^2 - q^2)^2  +4 m_B^2 (m_s^2 - s_0)}}{2 m_B^2}. $$

Our main goal is to represent the result for the 
invariant amplitude $\Pi^{(A1)}(q^2,p^2)$ in the form of dispersion relation (\ref{eq:OPEdispform}) in the variable $p^2$ at fixed $q^2<0$. Integrating $\mathcal{I}_{n, r}$ over loop momentum in (\ref{eq:PiA1res}), yields 
well-defined analytical structure in $p^2$.  More specifically, terms originating from the loop-momentum integrals
contain $1/\Delta_{A1}$ or  $\ln(\Delta_{A1})$, yielding, respectively, poles and logarithmic cuts at $p^2=s>0$, as seen from the transformed form of $\Delta_{A1}$ in 
(\ref{eq:Delta}). In their turn, the coefficients $\tilde{B}_i^{(j,\phi(\Phi))}$ have only polynomial dependence on $p^2$, hence in these coefficients   $p^2$ can be effectively replaced by $s$, anticipating Borel transform. The resulting OPE spectral density  $(1/\pi) {\rm Im}_{s}\, \Pi^{(A1)}(q^2,s)$ (see Appendix~\ref{app:imag_parts} for imaginary parts of different functions), is then inserted in the dispersion integral (\ref{eq:OPEdispform}). Applying the Borel transform of the latter with respect to $p^2$ and subtracting the continuum contribution above the effective threshold $s_0$ in accordance with duality approximation, leads to the final expression for the contribution of diagram $(A1)$
in the form ready to be inserted in LCSR (\ref{eq:Phf}):
\begin{align}
     I^{(A1)}(M^2, q^2)= &   \   \frac{-1}{(4\pi)^2} \frac{1}{ m_b^2} \Bigg\{  \int_0^{\hat{\omega}_0} d\hat{\omega}\, e^{\frac{- \tilde{s}}{M^2}} \,  \int_0^1 dx \ \Bigg[ \log\bigg(\frac{(x^2 - x)(1-\hat{\omega})}{\mu^2}\bigg) \Bigg( \frac{\tilde{B}_0^{(0,\phi)}}{(1- \hat{\omega})}  \phi_B^+(\omega) \nonumber \\[-8pt] & + \bigg( \frac{\tilde{B}_0^{(1,\Phi)} + \tilde{B}_0^{(2,\Phi)} }{(1- \hat{\omega})}  -  \frac{1}{M^2} \frac{\tilde{B}_0^{(3,\Phi)} }{(1- \hat{\omega})^2} + \frac{\tilde{B}_0^{'(3,\Phi)} }{(1- \hat{\omega})^2} + 2 \frac{\tilde{B}_1^{(3,\Phi)}}{(1- \hat{\omega})} (x^2 - x)   \nonumber \\[-8pt]
    & + 2 x \frac{\tilde{B}_1^{(4,\Phi)}}{(1- \hat{\omega})} \bigg) \frac{\Phi_B^\pm(\omega)}{2}\Bigg) -  x \frac{\tilde{B}_1^{(4,\Phi)}}{(1- \hat{\omega})} \frac{\Phi_B^\pm(\omega)}{2}  -\frac{1}{1 - x} \  \Big(- \frac{1}{M^2} \frac{\tilde{B}_0^{(4,\Phi)} }{(1- \hat{\omega})^2}  \nonumber \\[-8pt] &  + \frac{\tilde{B}_0^{'\ (4,\Phi)}}{(1 - \hat{\omega})^2}  \Big)  \frac{\Phi_B^\pm(\omega)}{2} +  \Big( \frac{-1}{M^2} \frac{\tilde{B}_0^{(3,\Phi)} }{(1- \hat{\omega})^2}  + \frac{\tilde{B}_0^{'\ (3,\Phi)}}{(1 - \hat{\omega})^2}  + \frac{\tilde{B}_1^{ (3,\Phi)}}{(1 - \hat{\omega})} (x^2-x)  \Big)  \frac{\Phi_B^\pm(\omega)}{2} \Bigg]_{s=\tilde{s}}\nonumber \\[-8pt]
    & + \int_{m_s^2}^{s_0} ds\, \int d\hat{\omega} \int_0^1 dx \Bigg[ \frac{1}{(1- \hat{\omega})} \frac{d}{ds} \bigg(\big(\tilde{B}_0^{(0,\phi)} \phi_B^+(\omega) + (\tilde{B}_0^{(1,\Phi)} + \tilde{B}_0^{(2,\Phi)} \nonumber \\[-8pt] &\quad + 2 \tilde{B}_1^{(3,\Phi)}(x^2 - x) +2 \tilde{B}_1^{(4,\Phi)} x) \frac{\Phi_B^\pm(\omega)}{2} \big)e^{\frac{-s}{M^2}}\bigg) \Theta(s-\tilde{s}) \log|s - \tilde{s}|  \nonumber\\[-8pt] 
    &  - 2 (x^2- x)\left(  \tilde{B}_1^{(0,\Phi)} \phi_B^+(\omega) + (\tilde{B}_1^{(1,\Phi)} + \tilde{B}_1^{(2,\Phi)} )\frac{\Phi_B^\pm(\omega)}{2} \right)  e^{\frac{-s}{M^2}} \Theta(s-\tilde{s})  \nonumber\\[-8pt] 
    & - \frac{1}{(1 -\hat{\omega})^2}\frac{d^2}{ds^2} (\tilde{B}_0^{(3,\Phi)} e^{\frac{-s}{M^2}} ) \log|s - \tilde{s}| \Theta(s - \tilde{s}) \frac{\Phi_B^\pm(\omega)}{2} \Bigg] \Bigg\}\,,
    \label{eq:factA1f}
\end{align}
where $\tilde{B}_i^{\prime\,(j,\Phi)}$ is the derivative of $\tilde{B}_i^{(j,\Phi)}$ with respect to the variable $s$. The remaining form-factor-like contributions from diagrams $(A2),(A3)$ and $(B1),(B2),(B3)$ yield analogous invariant amplitudes $I^{(A2)},I^{(A3)}$ and $I^{(B1)},I^{(B2)},I^{(B3)}$ respectively, as provided in Appendix~\ref{app:invfuncs}.
Their coefficient functions denoted, respectively, as 
$\tilde{C}_i^{(j,\phi(\Phi))}$, 
$\tilde{A}_i^{(j,\phi(\Phi))}$
and 
$\bar{A}_i^{(j,\phi(\Phi))}$, 
$\bar{C}_i^{(j,\phi(\Phi))}$, 
$\bar{B}_i^{(j,\phi(\Phi))}$ 
are given in the ancillary file.

As already explained in the previous section at a qualitative level,
the remaining diagrams $(A4)$ and $(B4)$ do not develop an imaginary part
in $p^2$ within the duality interval, characteristic for the kaon channel and therefore do not contribute to the LCSR (\ref{eq:Phf}). To substantiate this 
 statement with a direct computation, we consider e.g.  
 diagram~$(A4)$. After loop-momentum integration, the only propagator in this diagram which retains a dependence on the variable $p^{2}$, has the denominator equal to
\begin{align*}
  (\omega v - q)^2 &= \hat{\omega} \left(p^2 - \big((1- \hat{\omega})m_B^2 - \frac{1-\hat{\omega}}{\hat{\omega}} q^2\big) \right)\,. \nonumber 
\end{align*}
At  $q^2<0$, this expression vanishes, yielding a pole at $p^2 \gtrsim \mathcal{O}(m_B^2 + |q^2|)$ in the diagram expression. That pole  is indeed 
well above the duality interval for the kaon channel. 
The same reasoning applies to diagram~$(B4)$ and also to diagram $(D3)$
with soft gluon.

An important and nontrivial property of the form-factor-like diagrams is that their expressions  are complex valued at $q^{2}<0$. 
The imaginary part is revealed for the  diagrams $(A1)$, $(A2)$, $(A3)$, as well as $(B2)$. The emergence of  imaginary part at spacelike $q^2$ is a known feature of $b\to s\gamma^*$ diagrams at NLO calculated for the inclusive $b\to s\ell^{+}\ell^{-}$ process \cite{Asatryan:2001zw} (see a detailed discussion in \cite{Khodjamirian:2012rm}).
An interesting question which is beyond our scope is the interpretation
of these discontinuities in terms of intermediate on-shell hadronic states.
As a result, the dispersion relation (\ref{eq:OPEdispform}) should be understood as a linear combination of two separate dispersion relations, splitting the integrand ${\rm Im}_{p^2}\Pi^{(OPE)}(s,q^2)$ into a real and imaginary part.

\subsection{Hard-scattering diagrams}
\label{sec:nfhg}
We now turn to the hard-scattering contributions represented by diagrams $(C1)$--$(C4)$. These diagrams all contain hard gluon emitted from the chromomagnetic operator and absorbed  by the spectator quark.  
They  enter the sum rule at the same order in $\alpha_s$ as the form-factor-like diagrams discussed in Sec.~\ref{sec:factodigs}. 
 
As an explicit example, we consider diagram~$(C1)$, where the photon is emitted from the $b$-quark line. 
The corresponding contribution to the correlation function (\ref{eq:corr}) reads:

\begin{align}
    \Pi_{\nu\mu} ^{(C1)}(q, p) &= -i f_B\, m_B \left(\frac{\alpha_s C_F}{32\pi}\right) \, m_b \,C_{8g}\, Q_b  
    \int_0^\infty d\omega \int \frac{d^4 k}{(2\pi)^4}  \nonumber\\[-6pt] 
    &\quad \times \Bigg\{ \frac{{\rm Tr}\!\big[ \hat{\mathcal{C}}_{\nu\eta\mu} \, \gamma_5 \big] k^\eta}{\mathcal{D}_{C1}} \, \phi_B^+(\omega)
    + \frac{ {\rm Tr}\!\big[ \hat{\mathcal{C}}_{\nu\eta\mu} \, \gamma^\alpha\gamma_5 \big] k^\eta}{2\mathcal{D}_{C1}} \, 
      \overset{\leftarrow}{\partial}_\alpha \Phi_B^\pm(\omega) \Bigg\}\,,
      \nonumber \\[-6pt]     
     &\quad = 
     ip_\nu q_\mu\Pi^{(C1)}(p^2,q^2)+ ...\,,
    \label{eq:intC1}
\end{align}
where the Dirac structure $\hat{\mathcal{C}}$ is given by:
\begin{align*}
\hat{\mathcal{C}}_{\nu\eta\mu} &\equiv  \gamma_\rho (\slashed{k} + \omega \slashed{v})\, \gamma_\nu \gamma_5 \, (\slashed{k} + \omega \slashed{v} - \slashed{p}-m_s)\, \sigma_{\rho \eta} (1+\gamma_5)  (m_b \slashed{v} - \slashed{q} + m_b)\, \gamma_\mu (1+\slashed{v})\,,
\end{align*}
and the short-hand notation for the denominator  is introduced: 
\begin{equation*}
    \mathcal{D}_{C1} \equiv \big((m_b v - q)^2 - m_b^2\big) \, k^2 \, (k + \omega v)^2 \, (k + \omega v - p)^2\,.
\end{equation*}
After performing the loop-momentum  integration, transforming the invariant amplitude
$\Pi^{(C1)}(p^2,q^2)$
to a form of dispersion relation in $p^2$, followed by Borel transform and continuum subtraction, the expression we obtain for the contribution of the diagram
$(C1)$ to LCSR turns out rather cumbersome. We found the way to simplify that, applying a transformation of Feynman parameters,
$x_1 = X y_1$ and $x_2 = X(1-y_1)$, such that $x_1 + x_2 = X$ and $0 \leq X, y_1 \leq 1$. The corresponding Jacobian is $|J| = X$,  
and we define
\begin{align}
   %\Delta_{C1} &= X(\mathcal{W}_{C1}\, p^2 - \mathcal{X}_{C1}) \,,\nonumber\\ 
    \mathcal{W}_{C1} &= -\bar{y}_1 \, (\bar{\hat{\omega}}\bar{X} + X\, y_1 ) \, , \nonumber\\ 
    \mathcal{X}_{C1} & = -\hat{\omega} \bar{X} \left( (q^2 - m_B^2) \bar{y}_1  - m_B^2 \, \hat{\omega}\right)\,,
\end{align} 
where the barred variables are introduced as, e.g.,  $\bar{X} = 1-X$.

With these definitions, we obtain:
\begin{align}
    I^{(C1)}(M^2, q^2) = & - \int_{m_s^2}^{s_0} ds\ e^{-s/M^2} \int_0^\infty d\hat{\omega}\, \int_0^1 dX\, dy_1 \, \frac{1}{m_b^2}  \, 
    \Bigg\{ \Bigg[ D_{0}^{{(0,\phi)}} \frac{\delta(\mathcal{W}_{C1} s -\mathcal{X}_{C1})}{\mathcal{W}_{C1}} \nonumber \\[-6pt] & \qquad - 2X\  D_{1}^{{(0,\phi)}}\, \Theta(\mathcal{X}_{C1} - \mathcal{W}_{C1} s)  \Bigg]\phi_B^+(\omega)\,+ 
     \Bigg[\bigg( D_0^{(1,\Phi)} + D_0^{(2,\Phi)} \nonumber \\[-6pt] &+ \frac{ \big(y_1 D_0^{(3,\Phi)} + (1 - y_1) D_0^{(4,\Phi)} \big) }{M^2} + \big(y_1 D_0^{'(3,\Phi)} + (1 - y_1) D_0^{'(4,\Phi)} \big) \nonumber \\[-6pt] & + 2 X\ \big( y_1 D_1^{(3,\Phi)} + (1-y_1) D_1^{(4,\Phi)} \big) \bigg) \frac{\delta(\mathcal{W}_{C1} s - \mathcal{X}_{C1})}{\mathcal{W}_{C1}}   \nonumber \\[-6pt] & - 2 X \big( D_1^{(1,\Phi)} + D_1^{(2,\Phi)} \big) \Theta(\mathcal{X}_{C1} - \mathcal{W}_{C1} s)   \Bigg]\frac{\Phi_B^\pm(\omega)}{2}\Bigg\}\,,
\end{align}
where the $\delta$-function fixes the domain of the integration region, while the $\Theta$-function implements the continuum subtraction. 
The treatment of the $\delta$-function is standard and reads:
\begin{align}
    \int_0^1 dX\, f(X)\, \delta(\mathcal{W}_{C1} \, s -\mathcal{X}_{C1}) = \frac{f(X_0)}{\left(\frac{\partial}{\partial X}(\mathcal{W}_{C1}\, s - \mathcal{X}_{C1})\right) \big|_{X=X_0}} \, \Theta\left( 0 < X_0 < 1\right)\,.
\end{align}

The remaining hard-scattering contributions to LCSR: $I^{(C2)}$, $I^{(C3)}$, and $I^{(C4)}$, are obtained analogously.
Their explicit analytical forms are provided in Appendix~\ref{app:invfuncs}. The corresponding coefficient 
functions denoted as  $B_i^{(j,\phi(\Phi))}$, 
$A_i^{(j,\phi(\Phi))}$, and $C_i^{(j,\phi(\Phi))}$ 
for diagrams $(C2)$, $(C3)$, and $(C4)$, respectively are presented in the ancillary file.

We also note that, in contrast with the form-factor-like contributions, none of the hard-scattering diagrams $(C1)$--$(C4)$ develops an imaginary part in the spacelike region $q^{2}<0$.

\subsection{Soft-gluon diagrams}
\label{sec:softg}
In addition to the hard-gluon contributions discussed above, the correlation function (\ref{eq:corr}) involves  diagrams, in which a low-virtuality
gluon belonging to the $B$-meson 
long-distance structure is exchanged with the chromomagnetic operator $O_{8g}$.
These are diagrams $(D1)$--$(D3)$ described in terms of  $B$-meson three-particle DAs\,, convoluted with a 
short-distance kernel, which is represented by two off-shell quark propagators. The relevant definitions and the explicit parameterisations of the three-particle DAs are collected in Appendix~\ref{app:DAs}. 

We only have to evaluate diagrams $(D1)$ and $(D2)$, where the virtual photon is emitted, respectively, from the $b$-quark and $s$-quark lines. 
As discussed in Sec.~\ref{sec:factodigs}, diagram $(D3)$ with a photon emission from the spectator quark does not  contribute to the sum rule, since the corresponding pole in $p^2$ lies far away from the duality interval of the kaon channel.  

As an explicit example, we discuss a 
soft-gluon contribution from diagram~$(D1)$, corresponding to a virtual photon emission from the $b$-quark. This computation is naturally simpler than in the 
case of hard-gluon diagrams, due to absence of the loop. Contracting the $b$ and $s$-quark fields into propagators,  we decompose the 
vacuum-to-$B$ matrix element of the remaining $b$, $\bar{s}$ and gluon
fields into a combination of 
$B$-meson three-particle DAs. After performing the coordinate integration, the resulting invariant amplitude takes the form:
\begin{align}
    \Pi^{(D1)} (p^2, q^2) = & \sum_{n=1,2}\frac{f_B m_b}{8\pi^2} C_{8g} \int_0^\infty d\omega \int_0^\infty d\xi \frac{1}{[(\omega v -p)^2]^n} \Big[ F_n^{(\Psi_{AV})} (q^2,\omega)(\Psi_A(\omega, \xi ) - \Psi_V(\omega, \xi )) \nonumber \\& + F_n^{(\Psi_{V})}(q^2,\omega) \Psi_V(\omega, \xi ) + F_n^{(X_{A})}(q^2,\omega) \overline{X}_A(\omega, \xi ) + F_n^{(W Y_{A})}(q^2,\omega) \big(\overline{W} (\omega,\xi)  \nonumber\\ & + \overline{Y}_A(\omega, \xi)\big) + F_n^{(\tilde{X}_{A})}(q^2,\omega) \overline{\tilde{X}}_A(\omega, \xi)+ F_n^{(\tilde{Y}_{A})}(q^2,\omega) \overline{\tilde{Y}}_A(\omega, \xi)    \Big]\,.
\end{align}

The coefficient functions $F_n^{(X)}$, where $X$ stands for a three-particle DA, are given in Appendix~\ref{app:coeff}.
Our result supersedes the earlier calculation of the same diagrams in \cite{Khodjamirian:2012rm}, by including a complete set of three-particle DAs.
We also spotted a difference in one of the coefficients, (see Appendix~\ref{app:coeff} for details) which however has a negligible impact on the numerical results.

A transition from the above expression for the invariant amplitude
to a final integral over spectral density $I^{(D1)}(M^2, q^2)$, obtained after casting in a dispersion form, Borel transform and duality subtraction
is reduced to the following replacements:
\begin{align}
    \int_0^\infty d\omega \frac{f(q^2, \omega)}{(\omega v - p)^2} \to & - \int_0^{\hat{\omega}_0} \frac{d\hat{\omega}}{(1 - \hat{\omega})} e^{-\tilde{s}/M^2} f(q^2 , \omega)\,,\  \nonumber \\ 
    \int_0^\infty d\omega \frac{f(q^2, \omega)}{((\omega v - p)^2)^2} \to &  \int_0^{\hat{\omega}_0} \frac{d\hat{\omega}}{(1 - \hat{\omega})^2} e^{-\tilde{s}/M^2} \frac{f(q^2 , \omega)}{M^2} + \frac{e^{-\tilde{s}(\hat{\omega}_0)/M^2} f(q^2 , \hat{\omega}_0)}{m_B (1 - \hat{\omega}_0)^2 \left(1- \frac{q^2}{m_B^2(1 - \hat{\omega}_0)^2} \right)}\,. 
\end{align}
The remaining soft-gluon diagaram $(D2)$ is computed  in an analogous way.
Its final expresison is provided in Appendix~\ref{app:invfuncs}
and the associated coefficient functions denoted as 
$\tilde{F}_n^{(X)}$ are collected in the ancillary file.

\subsection{Power counting}
\label{sec:powercount}
Since we have at hand all contributions to the LCSR (\ref{eq:lcsr})  calculated within one and the same approach and input, it is instructive to investigate the hierarchy of these 
contributions with respect to the powers of inherent large scales.
In general, here we deal with the following parametric hierarchy of scales: 
\begin{equation}
\big\{m_b,m_B\big\}\gg \big\{ \sqrt{|q^2|},\mu , M\big\}\gg 
\big\{  \Lambda_{QCD}, \sqrt{s}_0,\bar\Lambda\big\}\equiv \Lambda\,,
\label{eq:scales} 
\end{equation}
where $\mu$ ($M$) is the renormalisation (Borel) scale and on r.h.s. we introduce a generic hadronic scale 
$\Lambda \lesssim 1$ GeV 
which spans from $\Lambda_{QCD}$ to $\bar{\Lambda}=m_B-m_b$. 

Within the LCSR framework, the non-perturbative input is provided by the $B$-meson DAs, depending on the momentum variable $\omega\sim\mathcal{O}(\Lambda)$. The latter condition is 
justified, since the realistic models of $B$-meson DAs are suppressed at large~$\omega$. Moreover, the characteristic soft scale 
of these distributions 
is governed by the inverse moment $\lambda_B$ of the leading-twist DA $\phi_B^+(\omega)$, defined in (\ref{eq:invmom}), and this parameter is also of ${\mathcal O}(\Lambda)$. 

Taking into account the scale hierarchy, discussed above, we analysed 
the relative role of separate diagrams in LCSR.
The contribution of each diagram to the invariant amplitude of the correlation function can be represented as a convolution of the perturbatively calculable hard kernel with the $B$-meson DA. Schematically, up to the common prefactor $(C_F/2)(f_B m_B/4)m_b$, 
these amplitudes take the form
\begin{align}
    \Pi^{(X)}\sim \int_0^\infty d\omega\,\phi_B^+(\omega)\,K^{(X)}(\omega,q^2)\;+\;(\Phi_B^\pm\text{-term})\,,
\end{align}
where  $(X)=(A1),... $ labels the individual diagrams. For $(D1)$ and $(D2)$, the two-particle DAs are replaced with three-particle DAs. 
Determined by the dependence of the kernels $K^{(X)}$
on the variables $\omega$ and $q^2$, 
the amplitudes $\Pi^{(X)}$ for each category of diagrams reveal distinctively different parametric values: 
\begin{itemize}
    \item for the form-factor-like diagrams ($A,B$), we obtain  $K(\omega)\!\propto\!1/\omega$, yielding
\begin{align}
    \;\Pi^{(A+B)} \;\sim\;\left(\frac{\alpha_s}{4\pi}\right)\frac{1}{\lambda_B}\Big[1 +\mathcal O\left(\frac{\Lambda}{m_B} \right)\Big]\,,
\end{align}

    \item for the hard-scattering diagrams  $(C)$, $K(\omega)\!\propto\!(1/\omega) \times \mathcal O(\Lambda/m_B) $, yielding
   \begin{align}
    \;\Pi^{(C)} \;\sim\;\left(\frac{\alpha_s}{4\pi}\right)\frac{1}{\lambda_B}\,\times \mathcal{O}\left(\frac{\Lambda}{m_B} \right) \,,
\end{align}

    \item  the soft-gluon diagrams ($D$) involve three-particle $B$-meson DAs with normalisation $\lambda_{E,H}^2\!\sim\!\mathcal O(\Lambda^2)$, leading to
\begin{align}
    \Pi^{(D)} \;\sim\; 
    \frac{1}{ \lambda_B }\times \mathcal{O}
    \left(\frac{\lambda_{E,H}^2}{m_B^2 }\right)
    \,.
\end{align}

\end{itemize}
Note that the power hierarchy of diagrams obtained above does not change after Borel transform and duality subtraction. We come to a conclusion that in the $O_{8g}$ contribution to the nonlocal form factor, the diagrams with form-factor-like topologies are parametrically leading while both hard-scattering and soft-gluon contributions are suppressed by at least one power of $\mathcal O\left(\Lambda/m_B \right)$. 
Moreover, albeit the soft-gluon contribution is suppressed by additional power of $\Lambda/m_B$ with respect to the hard-scattering contribution,
this extra power suppression can be compensated by  $\alpha_s/4\pi$ which is 
numerically of the same order.

\section{Numerical Results}
\label{sec:results}
We begin with the numerical evaluation of the LCSR in~(\ref{eq:lcsr}) in the region $q^{2}<0$. The input values of quark and meson masses, e.m. and strong couplings, together with the decay constants of $B$-meson and kaon are collected in Table~\ref{tab:inputvalues}.  
Since our calculation is performed at spacelike momentum transfer, we employ the $\overline{\text{MS}}$ scheme. The expressions of the $B$-meson DAs used in our analysis are given in Appendix~\ref{app:DAs}. They
are determined by two independent parameters
for which we adopt QCD sum rule estimates. In particular, for the 
inverse moment of the twist-2 DA, $\lambda_B$, we take the result of 
\cite{Braun:2003wx} (a close interval was obtained later in 
\cite{Khodjamirian:2020hob}).
For the ratio $R=\lambda_E^2/\lambda_H^2$
of the parameters characterising quark-antiquark-gluon  DAs, we use the average of two independent 
sum rule analyses \cite{Nishikawa:2011qk} and \cite{Rahimi:2020zzo} \footnote{This average is kindly provided by Aritra Biswas.}.

Furthermore, for the two specific parameters of LCSR (\ref{eq:lcsr}), the Borel mass $M$ and
effective threshold $s_0$ for the channel of the kaon interpolation current, it is natural to adopt the values in the same
ballpark as for the analogous LCSRs with $B$-meson DAs for the $B\to K$ local form factors. In the original analysis \cite{Khodjamirian:2006st} of the latter sum rule, the interval $M^2=1.0\pm 0.5$ GeV$^2$ was taken, together with the value $s_0=1.05$ GeV$^2$. These values were taken  from the earlier analysis \cite{Khodjamirian:2003xk}
of two-point sum rules for the kaon decay constant in the same channel of the
kaon interpolating current (\ref{eq:jkaon}). In the later update \cite{Gubernari:2018wyi} of the LCSR for the $B\to K$ form factors, the same interval for the Borel mass was taken and for $s_0$
a small uncertainty was included which we double. To stay on a  conservative side,
we also assume that the Borel mass and the effective  threshold are uncorrelated. 

To adopt an optimal scale for strong coupling and 
quark masses, we note that  
the typical virtuality of the hard kernel is in the ballpark of  $m_b\,\Lambda_{\rm QCD}$. Accordingly, we evaluate the scale-dependent amplitude ${\cal H}^{BK,O_8}(q^2)$ given by the 
LCSR (\ref{eq:lcsr})
at $\mu=2~\text{GeV}$. Varying the scale around this value, we find that noticeable scale dependence is revealed mainly, through the
form-factor-like contributions of diagrams $(A)$ and $(B)$, while the hard-scattering and soft-gluon terms are almost insensitive to~$\mu$. We also neglect
possible renormalisation effects of the DAs, in particular
the scale-dependence of the inverse moment $\lambda_B$, 
having in mind a large uncertainty of this quantity.

\begin{table}[t]
    \renewcommand{\arraystretch}{1.5}
    \centering
    % \begin{tabular}{|c|>{\centering\arraybackslash}p{7cm}|>{\centering\arraybackslash}p{6cm}|}
    \begin{tabular}{|c|c|c|}
         \hline
         Parameter & Value & Ref. \\ \hline \hline
         $m_{K^0}$ & $497.611\pm0.013$~MeV & \cite{ParticleDataGroup:2024cfk} \\ 
         $m_{B^0}$ & $5279.63\pm0.20$~MeV & \cite{ParticleDataGroup:2024cfk} \\ 
         $m_b$ & $4.183\pm0.004$~GeV & \cite{ParticleDataGroup:2024cfk} \\
         $m_c$ & $1.2730\pm0.0046$~GeV & \cite{ParticleDataGroup:2024cfk} \\
         $\alpha_{\rm em} $ & $ 1/128$ & \cite{ParticleDataGroup:2024cfk} \\ 
         $\alpha_s (m_Z)$ & $0.1180\pm0.0009$ & \cite{ParticleDataGroup:2024cfk} \\ 
         $f_B$ &  $190.0\pm 1.3$~MeV & \cite{FlavourLatticeAveragingGroupFLAG:2024oxs} \\
         $f_K$ &  $155.7 \pm 0.3$~MeV & \cite{FlavourLatticeAveragingGroupFLAG:2024oxs}\\
         $ \lambda_B$ & $460 \pm 110$~MeV & \cite{Braun:2003wx}\\
         % $\lambda_E $ & $   \textrm{MeV}$ & (may be considered ratio of R = lamE/lamH ) \\
         % $\lambda_H $ & $   \textrm{MeV}$ &  \\
         $ R (\equiv \lambda_E^2/\lambda_H^2)$ & $0.3^{+0.17}_{-0.12}$ & see text \\  
         $M^2$ & $1.0 \pm 0.5$ GeV$^2$ & see text  \\ 
         $s_0$ & $1.05 \pm 0.05$ GeV$^2$ & see text \\
        % $C_1(m_b)$ & $1.010$ &  \\
        % $C_2(m_b)$ & $-0.290$ &  \\
        % $C_{8g}(m_b)$ & $-0.171$ &  \\
         \hline
    \end{tabular}
    \caption{Input parameters used in the numerical evaluation of the LCSR (\ref{eq:lcsr}). }
    \label{tab:inputvalues}
\end{table}

According to our definition (\ref{eq:hme}), the nonlocal form factor $\mathcal{H}^{(BK,O_8)}(q^2)$ determined from  LCSR (\ref{eq:lcsr}) contains the Wilson coefficient  $C_{8g}$ as a factor. We consider the scale of this coefficient, as well as of the coefficients $C_{1,2}$  entering the ratio (\ref{eq:ratioO8}), to be independent of the scale we use to calculate the nonlocal form factor from LCSR. In our numerical analysis we use $C_1(m_b)=1.010$, $C_2(m_b)=-0.290$, $C_{8g}(m_b)=-0.171$~\cite{Mahmoudi:2008tp}.
 
\begin{figure}[t!]
	 \centering
\includegraphics[width=0.8\textwidth]{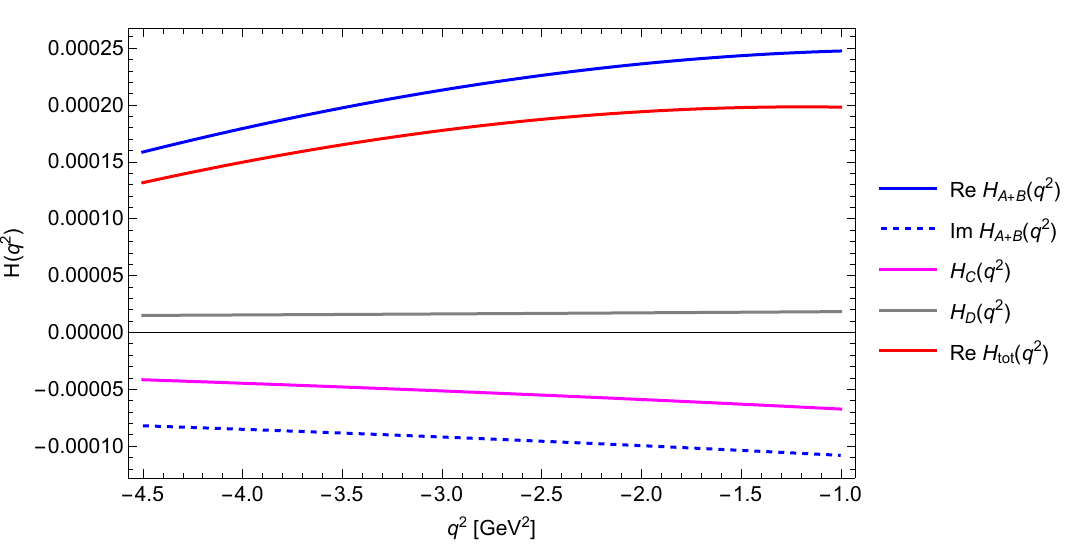}	
\caption{Individual contributions to the rescaled non-local form factor
for the operator $O_{8g}$, defined according to (\ref{eq:H}), in the spacelike region $q^{2}<0$. 
The blue and dashed-blue lines represent the real and imaginary parts of the form-factor-like contribution $H_{A+B}$, respectively. The magenta and gray curves correspond to the hard-scattering and soft-gluon contributions, $H_{C}$ and $H_{D}$. The total real part, $\mathrm{Re}\,H_{\text{tot}}(q^{2})$, obtained from their coherent sum, is shown in red.}
	 \label{fig:Htot_q2}
\end{figure}

For numerical illustration, it is more convenient to isolate the ``genuine" hadronic amplitude determined from LCSR  from this short-distance factor, introducing 
\begin{equation} 
H(q^2)\equiv \frac{\mathcal{H}^{(BK,O_8)}(q^2)}{C_{8g}}\,.
\label{eq:H}
\end{equation}
Figure~\ref{fig:Htot_q2} displays the real and imaginary parts of different contributions to this quantity,
according to the classification 
defined in Secs.~\ref{sec:factodigs}--\ref{sec:softg}. 
For consistency, in all diagrams, the $B$-meson DAs up to twist-3
are included.
   
The form-factor-like diagrams, labeled $(A,B)$, yield the dominant part of the nonlocal form factor, while the hard-scattering $(C)$ and soft-gluon $(D)$ diagrams are suppressed by approximately one order of magnitude, in agreement with the power-counting analysis presented in Sec.~\ref{sec:powercount}.  
The imaginary part of $(A+B)$ originates entirely from the logarithmic branch cuts of the hard kernel and remains small throughout the spacelike region.
Notice that the nonlocal contributions from $O_{8g}$ exhibit only a mild $q^{2}$ dependence across the considered spacelike region. 

\begin{figure}[h!]
	 \centering
\includegraphics[width=0.8\textwidth]{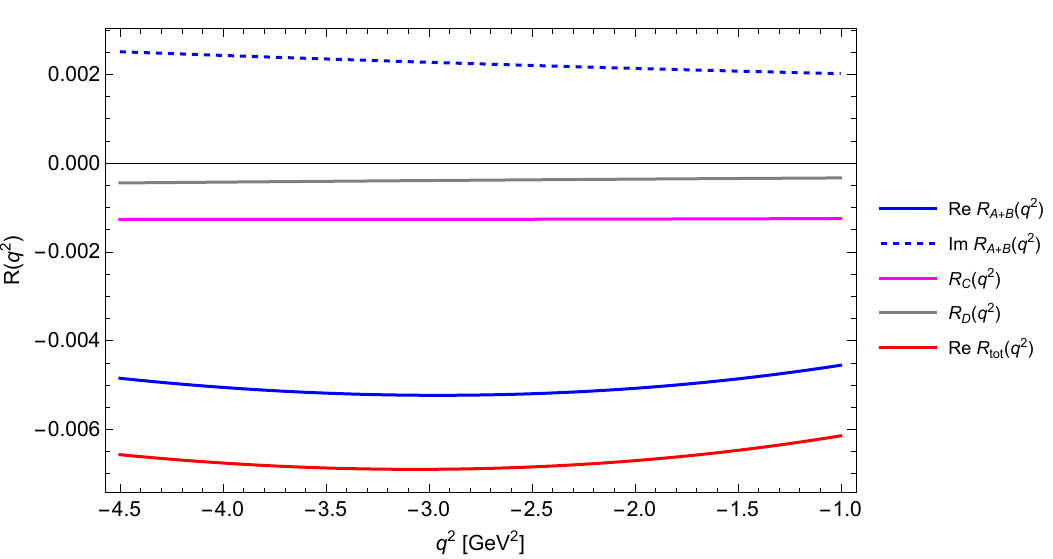}	
\caption{The blue (solid and dashed) lines show the real and imaginary parts of the factorisable term, while the magenta and gray curves represent the nonfactorisable hard- and soft-gluon pieces. The total ratio $\mathrm{Re}\,R_{\text{tot}}(q^{2})$ is shown in red. }
	 \label{fig:plotR}
\end{figure}
     
Furthermore, Figure~\ref{fig:plotR} displays the ratio (\ref{eq:ratioO8}) in the spacelike $q^2$ region. Here, we take 
the prediction of LCSR for $B\to K$ form factor at negative $q^2$  from \cite{Gubernari:2018wyi}, where the full higher-twist contributions for three-particle DAs are included. This choice is consistent with the approximation adopted in our calculation, and in accordance with using the same Borel interval and threshold in the kaon channel. The real part of the ratio is negative and dominated by the form-factor-like term $R_{A+B}$ (blue), while the hard-scattering and soft-gluon pieces $R_{C}$ (magenta) and $R_{D}$ (gray) are suppressed. 
The imaginary part (dashed blue), arising from the logarithmic branch cuts, remains small. The overall magnitude of $|R_{\text{tot}}(q^{2})| \lesssim 6\times10^{-3}$ 
indicates that the $O_{8g}$ contribution is strongly suppressed compared to the $O_{1,2}^{c}$ term. Importantly, in this suppression the ratio of Wilson
coefficients acts in the opposite direction, since the ratio (\ref{eq:ratioO8})
contains a factor $C_{8g}/(C_1/3+C_2)>1$.

For the soft-gluon contributions, we also studied the effect of additional three-particle $B$-meson DAs not included in \cite{Khodjamirian:2012rm} and found that the numerical impact of this modification is inessential.
 
\begin{table}[t]
\begin{center}
\setlength\extrarowheight{3pt}
% \hspace*{-1.cm}
\scalebox{0.96}{
%%%%%%%%%%%%%%%%%%%%%%%%%%%%%%%%%%%%%%%%%%%%%%%%%%%%%%%%%%%%%%%%%%%%%%%%%%
\begin{tabular}{|c||c|c|c|c|}
\hline
\multicolumn{5}{|c|}{Nonlocal form factor ${H}^{(BK,O_8)}$ and its components at $q^2=-4.0\text{ GeV}^2$ (in units of $10^{-4}$)}\\\hline
$\gamma^*$ & form-factor-like ($A$) & form-factor-like ($B$) & hard-scattering ($C$)  & soft-gluon ($D$)  \\\hline
% $\gamma^*$ & \multicolumn{2}{c|}{hard-fact}  & hard-nonfact  & soft-nonfact  \\
% & A & B & C & D\\\hline
(1) & $0.046 + i\ 0.014 $ & $0.436$ & $\phantom{+}0.205$ & $\phantom{+}0.297$ \\
(2) & $0.446 - i\ 0.623$  & $0.482 - i\ 0.272 $ & $-0.180$ & $-0.150$ \\
(3) & $0.007 - i\ 0.04$ & $0.562$ & $-0.409$ & 0 \\
(4) & 0 & 0 & $-0.099$ & -- \\\hline\hline
subtotal & \multicolumn{2}{l|}{~~~~~~$1.979(0.146) -i\ 0.885(0.155)$} & \multicolumn{1}{l|}{$ -0.482(0.251)$} & \multicolumn{1}{l|}{$0.151(0.024)$}\\\hline\hline
total & \multicolumn{4}{c|}{~~~~~~$1.648(0.291) -i\ 0.885(0.155)$} \\\hline

%%%%%%%%%%%%%%%%%
\end{tabular}
%%%%%%%%%%%%%%%%%%%%%%%%%%%%%%%%%%%%%%%%%%%%%%%%%%%%%%%%%%%%%%%%%%%%%%%%%%
}
\caption{ The entries~(1)--(4) in the first column correspond to the different photon emission topologies illustrated in Fig.~\ref{fig:O8gdiag}. The columns list the separate form-factor-like~($A,B$), hard-scattering~($C$), and soft-gluon~($D$) components.
}
\label{tab:numerical_results} 
\end{center} 
\end{table}

As another numerical illustration, in Table~\ref{tab:numerical_results} we present our results for the nonlocal form factor generated by the 
$O_{8g}$ operator at a momentum transfer squared $q^2=-4.0\text{ GeV}^2$. As previously found, the spectator-emission configurations ($A4$) and ($B4$) do not develop spectral density dual to the kaon pole in the 
dispersion relation and therefore vanish, as indicated 
in Table~\ref{tab:numerical_results}.

To obtain the uncertainties quoted in this table, we used a Monte Carlo approach, sampling the input parameters $M^2$, $s_0$, $\lambda_B$, $R$, and $f_B$ as independent variables. For $\lambda_B$, $R$, and $f_B$ we employ Gaussian distributions with central values and standard deviations taken from Table~\ref{tab:inputvalues}. For the Borel parameter $M^2$ and the effective threshold $s_0$, truncated Gaussian distributions are used to avoid going beyond the validity of the permitted Borel window. The error for each of the twelve subdiagrams is computed individually; the resulting errors are then added in quadrature to obtain the uncertainty for each diagram class, and finally for the total result.

It is instructive to compare our new results  for the hard-gluon contributions with the existing results in the literature obtained in the framework of QCDf. However, one has to keep in mind that such a comparison is rather qualitative, because our diagrammatic classification in the LCSR framework does not have a one‑to‑one correspondence with 
QCDf diagrams.
 
In particular, we compare the form-factor-like contributions 
(the sum of diagrams $(A)$ and $(B)$) obtained here, with the similar contribution to QCDf at $q^2<0$, using the hard-gluon $b\to s \gamma^*$ kernel calculated at NLO in \cite{Asatryan:2001zw} and cast into a form of the coefficient function $F_8^{7,9}$. This kernel is then factorised 
with the local $B\to K$ form factor (see \cite{Khodjamirian:2012rm}\,, where
this contribution was estimated at $q^2<0$). Using the analytical expression for the
function $F_8^{7,9}$ and  the local $B\to K$ form factor from~\cite{Gubernari:2018wyi}, we obtain, at the central values of the input:
$${H}^{(BK,O_8)}_{(A+B)}\Big|_{QCDf}= \big(1.39 - i\,0.84 \big)\times 10^{-4}\,,$$ in the same ballpark as
our numerical estimate from LCSR presented in Table~\ref{tab:numerical_results}.

For the hard-scattering contributions, our LCSR includes all four diagrams, $(C1)$--$(C4)$.
In the original QCDf analysis ~\cite{Beneke:2001at}, only the diagram analogous to $(C4)$ is taken into account, 
and the remaining diagrams  are not included, being power-suppressed, according to the power-counting 
in QCDf. In contrast, in LCSR, the contributions of all four hard-scattering diagrams are of the same order.
On the other hand, we find a strong numerical cancellation between the diagrams $(C1)$ and $(C2)$. This effect is caused 
by opposite signs of these diagrams, where in $(C_1)$ (($C_2))$ the virtual photon is emitted from the incoming $b$-quark (outgoing $s$-quark).
After this cancellation, the resulting hard-scattering contribution presented in 
Table~\ref{tab:numerical_results}
is smaller than the form-factor-like one. Our result has the same sign and a compatible magnitude 
with the estimate 
$${H}^{(BK,O_8)}_{(C)}\Big|_{QCDf}= -0.78\times 10^{-4}\,,$$ 
that we obtain using the formula for the hard-scattering diagram
in QCDf derived in~\cite{Beneke:2001at}.  

For the soft-gluon contributions given by diagrams $(D1)$--$(D2)$, it is important to compare our result presented in Table~\ref{tab:numerical_results}
with the previous calculation done within the same LCSR method in ~\cite{Khodjamirian:2012rm}.
The difference between these two calculations are: (i) a slight 
difference in the formula for the coefficient $F_1^{(X_A)}$ in the LCSR, and  (ii) 
a more complete set of $B$-meson three-particle DAs employed here and taken from 
\cite{Braun:2017liq}. Using the expression presented in~\cite{Khodjamirian:2012rm}
we recalculated the soft-gluon contribution and found:
$${H}^{(BK,O_8)}_{(D)}\Big|_{LCSR[7]}=0.159 \times 10^{-4}\,,$$
very close to our current estimate. This comparison indicates that both differences mentioned above are irrelevant for the numerical results.
In fact, the $B$-meson three-particle DAs of higher than three  twists, have an insignificant 
impact on the numerical result for the soft-gluon contribution. 

Note that the numerical analysis carried out here does not take into account
renormalisation effects related to the  $O_{8g}$ operator, such as the scale-dependence  
or mixing with four-quark operators. These effects will become relevant when contributions of 
other effective operators to the nonlocal form factor will be computed 
in a future work.

\section{Conclusion and outlook}
\label{sec:discussions}

In this paper, for the first time, we demonstrated that it is possible 
to compute the nonlocal form factor
in $B\to K \ell^+\ell^- $ at spacelike momentum transfer $q^2<0$, applying  the 
method of QCD LCSRs with $B$-meson DAs. Specifically, we obtained 
the contribution of the chromomagnetic penguin operator $O_{8g}$,
a task that  is technically less involved, as compared to the case of 
current-current operators, which demands computation of two-loop diagrams. 
Our main results are the  contributions to the nonlocal form factor generated by hard-gluon 
exchanges between quarks in this hadronic matrix element. We also recalculated and slightly updated   the previously known soft-gluon contributions.
Importantly, the same LCSR approach is applicable to the local $B\to K$
form factors equally important for  $B\to K \ell^+\ell^- $ decays.

We identify three different categories of diagrams contributing to LCSR, two of them
with hard-gluon exchanges (the form-factor-like and hard-scattering diagrams)
and the third category involving the soft-gluon effects. We computed and presented  analytical expressions for the spectral density of all diagrams
entering the  LCSR. Our results confirm that the nonlocal form factor
is a complex-valued quantity already at $q^2<0$, a property that can be interpreted as
the presence of intermediate on-shell light-hadron states. We also established
the power-counting scheme of different contributions, revealing the dominance of
form-factor-like contributions.

Our numerical analysis shows that, compared to leading-order charm-loop contribution, the nonlocal effect in $B\to K \ell^+\ell^- $ 
due to the chromomagnetic penguin operator is at sub-percentage level in the spacelike region. Furthermore, comparison with the previous calculations within QCDf framework indicates a reasonable agreement.

Our future task planned for the next work in this direction is a LCSR calculation 
of the contributions generated by the leading $O_{1,2}^c$ operators and similar four-quark operators.
We will extensively use the computational routine developed in this paper.
While principal elements of the method are already worked out here, there will be additional 
difficulties of both technical and conceptual origin. First of all, one will deal with two-loop diagrams, including the primary charm-loop. Secondly, concerning soft-gluon effects, one will encounter a problem of a proper
light-cone expansion of the correlation function, having in mind that there are three different points of quark-antiquark-gluon emission. 
This problem was recently discussed in detail in 
\cite{Piscopo:2023opf} in a context of a  LCSR calculation of the $B\to D\pi$ nonleptonic decays.

We believe that the results of this work bring us one step closer to resolving the problem 
of nonlocal effects  in the $b\to s$ FCNC transitions, providing an established QCD-based approach of LCSRs to analyse in detail this very intricate hadronic phenomenon.

\section*{Acknowledgements}
We thank Alexandre~Carvunis for discussions and involvement in the early stage of this project. 
D.M. is grateful to Thorsten~Feldmann for useful comments. \\
The research of A.K. was supported by the German Research Foundation (DFG) under grant 396021762 - TRR 257 ``Particle Physics Phenomenology after the Higgs Discovery''. A.K. and T.H. also acknowledge the hospitality during his visit to the Institut de Physique des 2 Infinis de Lyon, when this work was initiated.
T.H. is supported by the Cluster of Excellence ``Precision Physics, Fundamental Interactions, and Structure of Matter'' (PRISMA+ EXC 2118/1) funded by the German Research Foundation (DFG) within the German Excellence Strategy (Project ID 390831469). T.H. also thanks the CERN theory group for its hospitality during his regular visits to CERN where part of the work was done.
This research is funded in part by the National Research Agency
(ANR) under project no. ANR-21-CE31-0002-01.

\clearpage
\appendix

\section{Notations and conventions}\label{app:notation}
We consider the $|\Delta B | = 1 $ effective Hamiltonian~\cite{Grinstein:1988me,Misiak:1992bc,Buras:1994dj,Buchalla:1995vs}:
\begin{align}
\mathcal{H}_{\rm eff} = -\frac{4G_{F}}{\sqrt{2}} V_{tb}V_{ts}^* \sum_{i=1}^{10} C_i(\mu) O_i (\mu)\, + \text{h.c.}\,,
\label{eq:Heff}
\end{align}
with the operators:

\begin{eqnarray}
O_1^c& = &(\bar{s}_L \gamma_\mu c_L)(\bar{c}_L \gamma^\mu b_L)\,, \nonumber\\
     O_2^c& = &(\bar{s}_L^j \gamma_\mu c_L^i)(\bar{c}_L^i \gamma^\mu b_L^j)\,,
\end{eqnarray}  
where $q_{L(R)} = \frac{1-(+) \gamma_5}{2} q$ and $P_{L(R)} = \frac{1-(+) \gamma_5}{2}$.
The chromomagnetic operator $O_{8g}$, which 
plays the main role in this paper is  
already presented in (\ref{eq:C8g}). All other
effective operators entering (\ref{eq:Heff}) are not relevant here.

The sign convention adopted in this paper corresponds to the covariant derivative acting on quark field: $iD_\mu=i\partial_\mu +eQ_q A_\mu +g_sT^aA^a_\mu$, where $Q_q$ is the quark charge. We also use 
$\sigma_{\mu\nu}=i/2[\gamma^\mu,\gamma^\nu]$ 
and the  convention for the Levi-Civita tensor such that 
$\mbox{Tr}\{\gamma^\mu\gamma^\nu\gamma^\rho \gamma^\lambda\gamma^5\}
= 4i\epsilon^{\mu\nu\rho\lambda}$$,\epsilon^{0123}=-1$.

\section{\texorpdfstring{$B$}{B}-meson distribution amplitudes}
\label{app:DAs}
In HQET, the nonperturbative structure of the $B$-meson is encoded in a set of light-cone DAs, which are defined through matrix elements of nonlocal quark-gluon operators. We collect here the conventions and expressions used in our calculations, following~\cite{Braun:2017liq,Beneke:2018wjp}.

The two-particle (quark-antiquark) LCDAs are defined through
\begin{align}
\langle 0 \mid \bar{q}_2^\alpha(x)\,h_v^\beta(0) \mid \bar{B}_{q_2}(v) \rangle ~=&~
-\,\frac{i\,f_B\,m_B}{4}\,\int_{0}^{\infty} d\omega \, e^{-\,i\,\omega\,v\cdot x}\;\Bigl\{\,
(1 + \slashed{v})\Bigl[\,\phi_{+}(\omega)
~-\;g_{+}(\omega)\,\partial_\sigma\,\partial^\sigma
 \nonumber\\[-8pt] &+\;\Bigl(
\frac{\Phi_\pm(\omega)}{2}~-\;\frac{\bar{g}(\omega)}{2}\,\partial_\sigma\,\partial^\sigma
\Bigr)\,\gamma^\mu\,\partial_\mu\Bigr]\,
\gamma_5\Bigr\}^{\beta \alpha}\,,
\label{eq:2pLCDAs}
\end{align}
where $h_v$ is the heavy-quark field in HQET, and $\omega$ represents the light-cone projection of the spectator quark momentum.

The corresponding three-particle (quark-antiquark-gluon) LCDAs are defined through
\begin{align}
\langle 0 | \bar{q}_\alpha(x)\,G_{\mu\nu}(u\,x)\,h_{v\,\beta(0)} 
| \bar{B}(v)\rangle 
&= \frac{f_B\,m_B}{4} \int_{0}^{\infty} d\omega
\int_{0}^{\infty} d\xi\,
e^{-\,i\,\bigl(\omega + u\,\xi\bigr)\,v\cdot x} \nonumber\\[-8pt] 
& \Bigl\{\,(1 + \slashed{v})\Bigl[\,\bigl(v_\mu\,\gamma_\nu -\;  v_\nu\,\gamma_\mu\bigr)\,\bigl(\psi_{A} - \psi_{V}\bigr) - i\,\sigma_{\mu\nu}\,\psi_{V}  \nonumber\\[-8pt] 
&+ 
\bigl(\partial_\mu\,v_\nu - \partial_\nu\,v_\mu\bigr)\,\overline{X}_{A}-\;
\bigl(\partial_\mu\,\gamma_\nu - \partial_\nu\,\gamma_\mu\bigr)\,
\bigl[\,\overline{W} + \overline{Y}_{A}\bigr] \nonumber\\[-8pt] & +
i\,\epsilon_{\mu\nu\alpha\beta}\,\partial^\alpha\,v^\beta\,\gamma_{5} \overline{\tilde{X}}_A \, - i\,\epsilon_{\mu\nu\alpha\beta}\,\partial^\alpha\,\gamma^\beta\,\gamma_{5} \overline{\tilde{Y}}_A \, 
-\; u\,\bigl(\partial_\mu\,v_\nu \nonumber\\[-8pt] &-\partial_\nu\,v_\mu\bigr)\,\slashed{\partial}\,\overline{\overline{\overline{W}}}
+\; u\,\bigl(\partial_\mu\,\gamma_\nu - \partial_\nu\,\gamma_\mu\bigr)\,\slashed{\partial}\,\overline{\overline{Z}}\Bigr] \gamma_5
\Bigr\}_{\beta \alpha}\!(\omega,\xi)
\,,
\label{eq:3pDAs}
\end{align}
where $G_{\mu\nu}$ denotes the gluon field-strength tensor.  
The shorthand notations for the integrated DAs are
\begin{align}
    \overline{\psi}_{3 \rm p} (\omega, \xi) & \equiv \int_0^\omega d \tau \, \psi_{3 \rm p} (\tau, \xi)\,, \hspace{2cm}
    \overline{\overline{\psi}}_{3 \rm p} (\omega, \xi)  \equiv \int_0^\omega d \tau\, \int_0^\xi d\delta \, \psi_{3 \rm p} (\tau, \delta)\,.
\end{align}
To facilitate the power expansion in twist, the above amplitudes can be rewritten in terms of the conventional basis with definite twist components:
\begin{align}
   \psi_A(\omega, \xi) &=   \frac{1}{2} \left(\phi_3(\omega,\xi) + \phi_4(\omega,\xi)\right)\,, \nonumber\\[-8pt]
   \psi_V(\omega, \xi) &=   \frac{1}{2} \left(-\phi_3(\omega,\xi) + \phi_4(\omega,\xi)\right)\,,\nonumber\\[-8pt]
    X_A(\omega, \xi) &=   \frac{1}{2} \left(-\phi_3(\omega,\xi) - \phi_4(\omega,\xi) + 2 \psi_4(\omega, \xi)\right)\,,\nonumber \\[-8pt]
    Y_A(\omega, \xi) &=   \frac{1}{2} \left(-\phi_3(\omega,\xi) - \phi_4(\omega,\xi) + \psi_4(\omega,\xi) - \psi_5(\omega,\xi) \right)\,,\nonumber\\[-8pt]
    \tilde{X}_A(\omega, \xi) &=   \frac{1}{2} \left(-\phi_3(\omega,\xi) + \phi_4(\omega,\xi) - 2 \tilde{\psi}_4(\omega, \xi)\right)\,,\nonumber \\[-8pt]
    \tilde{Y}_A(\omega, \xi) &=   \frac{1}{2} \left(-\phi_3(\omega,\xi) + \phi_4(\omega,\xi) - \tilde{\psi}_4(\omega,\xi) + \tilde{\psi}_5(\omega,\xi) \right)\,,\nonumber \\[-8pt]
    W(\omega,\xi) &=  \frac{1}{2} \left(\phi_4(\omega,\xi) - \psi_4(\omega,\xi) - \tilde{\psi}_4(\omega,\xi) + \phi_5(\omega,\xi) + \psi_5(\omega,\xi) + \tilde{\psi}_5(\omega,\xi) \right),
\end{align}
which are obtained by inverting the relations in~\cite{Braun:2017liq}.
\subsection*{Exponential Model}
For the numerical analysis, we adopt the standard exponential model\cite{Grozin:1996pq,Braun:2017liq} for the $B$-meson LCDAs, defined as:
\begin{align}
    \phi^+_B(\omega) = & \frac{\omega}{\omega_0^2} e^{-\frac{\omega}{\omega_0}}\,,   \hspace{1cm}
    \phi^-_B(\omega) =  \frac{1}{\omega_0} e^{-\frac{\omega}{\omega_0}}\,,  \hspace{1cm} 
    \phi_3(\omega,\xi) =  \frac{R - 1}{(2/3)^3 (1+2R)^2\omega_0} \omega \xi^2\, e^{-\frac{(\omega +\xi)}{\omega_0}}\,, \nonumber \\
    \phi_4(\omega,\xi) = & \frac{R - 1}{(2/3)^3 (1+2R)^2} \xi^2\, e^{-\frac{(\omega +\xi)}{\omega_0}}\,, \hspace{0.8cm}
    \psi_4(\omega,\xi) =  \frac{R}{(2/3)(1+2R)}\ \omega \xi\, e^{-\frac{(\omega +\xi)}{\omega_0}}\,, \nonumber \\
    \tilde{\psi}_4(\omega,\xi,\mu_0) =  &\frac{1}{(2/3)(1+2R)} \omega \xi\ e^{-\frac{(\omega +\xi)}{\omega_0}}\,, \hspace{0.8cm}
    \psi_5(\omega,\xi,\mu_0) = - \frac{R}{(2/3) (1 + 2 R) \omega_0} \xi\, e^{-\frac{(\omega +\xi)}{\omega_0}}\,, \nonumber \\
    \tilde{\psi}_5(\omega,\xi,\mu_0) = &- \frac{1}{(2/3) (1 + 2 R) \omega_0} \xi\, e^{-\frac{(\omega +\xi)}{\omega_0}}\,, \hspace{0.8cm}
    \phi_5(\omega,\xi,\mu_0) =  \frac{R + 1}{(2/3)(1 + 2 R) \omega_0} \omega\, e^{-\frac{(\omega +\xi)}{\omega_0}}\,,
\end{align}
where $\omega_0$ in this model is equal to the inverse moment $\lambda_B$ of the twist-2 DA defined  as:
\begin{equation}
~~\int_0^\infty \! \!\!d\omega \,\frac{\phi_B^+(\omega)}{\omega} =\frac{1}{\lambda_B}\,,
\label{eq:invmom}
\end{equation}
and $ R \equiv \lambda_E^2 / \lambda_H^2$ denotes the ratio of the chromoelectric to chromomagnetic HQET parameters. This model provides a consistent description of the two- and three-particle $B$-meson DAs.

\section{Imaginary parts of different functions}
\label{app:imag_parts}
For convenience, here we collect expressions for the imaginary parts of the functions that appear in the dispersion relations of the invariant amplitudes. These equations are:
\begin{align*}
{\rm Im}\Big\{ \frac{1}{s - a + i\epsilon} \Big\} &= - \pi\, \delta(s-a)\,,\\[-8pt]
{\rm Im}\Big\{ \frac{1}{(s - a + i\epsilon)^n} \Big\} &=  \frac{(-1)^{n}}{(n-1)!} \, \pi\, \delta^{(n-1)}(s-a)\,,\nonumber \\[-8pt]
{\rm Im}\Big\{ \log(s - a+ i\epsilon) \Big\} &=  \pi\, \theta(a-s)\,,\nonumber \\[-8pt]
{\rm Im}\Big\{ \frac{\log(s - a+ i\epsilon)}{s - a+ i\epsilon}\Big\}  &=  \frac{d}{d s}(\pi \log|s - a| \Theta(a - s))\,, \nonumber \\[-8pt]
{\rm Im}\Big\{ \frac{\log(s - b + i\epsilon )}{s - a+ i\epsilon} \Big\} &=  -\pi \left( \log|s - b| \, \delta(s-a) - \mathcal{P}\left(\frac{1}{s-a}\right) \, \theta(b-s) \right)\,,\nonumber \\[-8pt]
   {\rm Im}\Big\{ \frac{\log(s - a+ i\epsilon)}{(s-a + i\epsilon)^2}\Big\} 
    &=  \pi\left( \delta'(s - a) - \frac{d^2}{d s^2}( \log|s-a| \Theta(a - s)\right)\,, \nonumber \\[-8pt]
{\rm Im}\Big\{ \frac{\log(s - b + i\epsilon)}{(s - a + i\epsilon)^2} \Big\} &=  \pi \left( \mathcal{P}\left(\frac{1}{(s-a)^2}\right) \, \theta(b - s) + \log|s - b| \, \delta'(s-a)  \right)\,.\nonumber
\end{align*}
The plus-type prescriptions entering the continuum subtraction are defined as:
\begin{align}
  \int_{m_s^2}^{s_0} ds\  \bigg[ \frac{1}{s - a}\bigg]_+ g(s) = &\ \int_{m_s^2}^{s_0} ds \frac{1}{ s- a }\ (g(s) - g(a)) \, \Theta(m_s^2 < a < s_0)\,, \nonumber \\[-8pt]
    \int_{m_s^2}^{s_0} ds \bigg[ \frac{1}{(s-a)^2}\bigg]_{++} g(s) =& \int_{m_s^2}^{s_0} ds \frac{g(s) - g(a) - (s - a) g'(a)}{(s - a)^2}\, \Theta(m_s^2 < a < s_0)\,.\nonumber
\end{align}
These relations are employed when constructing the spectral densities ${\rm Im}_s\, I^{(X)}(q^2,s)$ from the loop amplitudes and when implementing the continuum subtraction following the quark-hadron duality ansatz.

\section{Expressions for invariant functions}
\label{app:invfuncs}
In this appendix we provide the explicit expressions for the invariant functions $I^{(Ai)}$, $I^{(Bj)}$, and $I^{(Ck)}$ that arise from the diagrams shown in Fig.~\ref{fig:O8gdiag}. These results complement the main text, where only representative examples are displayed. 
Throughout this appendix, whenever a coefficient function depends on the variable~$s$, it is implicitly evaluated at the same kinematic parameter that appears in the Borel exponential that multiplies it. The relevant coefficients appearing in the invariant functions are collected in Appendix~\ref{app:coeff}. A prime on the coefficients indicates a derivative with respect to the dispersion
variable~$s$.

\subsection{Form-factor-like contributions}
For the form-factor-like diagrams $(A,B)$, the explicit form of 
$I^{(A1)}$ is given in the main text. The remaining contributions read:
\allowdisplaybreaks{
\begin{align}
    I^{(A2)}(M^2, q^2) =&\  \frac{1}{(4 \pi)^2}  \Bigg\{ \int_0^{\hat{\omega}_0}  \frac{d\hat{\omega}}{ (1- \hat{\omega})} \int dX \int dy_1  \frac{1}{a_s\ (\tilde{s} - s'_s)}  \bigg[ e^{-\tilde{s}/M^2} \bigg(\tilde{C}_0^{(0,\ \phi) }\phi_B^+(\omega) \nonumber \\[-8pt] & \qquad + \big(\tilde{C}_0^{(1,\ \Phi)} +\tilde{C}_0^{(2,\ \Phi)}+\tilde{C}_0^{(3,\ \Phi)}  + 2X (y_1 \tilde{C}_1^{(5,\ \Phi)} + (1 - y_1) \tilde{C}_1^{(6,\ \Phi)})\big) \frac{\Phi_B^\pm(\omega)}{2} \bigg)   \nonumber \\[-8pt] 
    &\qquad \times \Theta(s_0 - \tilde{s}) - e^{-s'_s/M^2}   \bigg(\tilde{C}_0^{(0,\ \phi) }\phi_B^+(\omega)  + \big(\tilde{C}_0^{(1,\ \Phi)} +\tilde{C}_0^{(2,\ \Phi)}+\tilde{C}_0^{(3,\ \Phi)}  \nonumber \\[-8pt] & \qquad + 2X (y_1 \tilde{C}_1^{(5,\ \Phi)}  + (1 - y_1) \tilde{C}_1^{(6,\ \Phi)})\big) \frac{\Phi_B^\pm(\omega)}{2} \bigg) \Theta(s_0 - s'_s) \bigg]  \nonumber \\[-8pt] 
    & +  \ \int_0^{\hat{\omega}_0}  \frac{d\hat{\omega}}{ (1- \hat{\omega})} \int dX \int dy_1 2 X  ( \log(\tilde{a}_s\ X (\tilde{s} - s'_s) )e^{-\tilde{s}/M^2}\big( \tilde{C}_1^{(0,\ \phi)} \phi_B^+(\omega) \nonumber \\[-8pt] & \qquad + (\tilde{C}_1^{(1,\ \Phi)} + \tilde{C}_1^{(2,\ \Phi)}+\tilde{C}_1^{(3,\ \Phi)} + 3 X (y_1 \tilde{C}_2^{(5,\ \Phi)} + (1 - y_1) \tilde{C}_2^{(6,\ \Phi)} )) \frac{\Phi_B^\pm(\omega)}{2} \big)  \nonumber \\[-8pt] 
    &  -  \int_0^{s_0} ds \int  \frac{d\hat{\omega}}{ (1- \hat{\omega})} \int dX \int dy_1 2 X  \bigg[\frac{1}{s - \tilde{s}}\bigg]_+ \Theta(s'_s-s) e^{-s/M^2} \bigg( \tilde{C}_1^{(0,\ \phi)} \phi_B^+(\omega) \nonumber \\[-8pt] & \qquad +  (\tilde{C}_1^{(1,\ \Phi)} + \tilde{C}_1^{(2,\ \Phi)}+\tilde{C}_1^{(3,\ \Phi)} + 3 X (y_1 \tilde{C}_2^{(5,\ \Phi)} + (1 - y_1) \tilde{C}_2^{(6,\ \Phi)} )) \frac{\Phi_B^\pm(\omega)}{2}  \bigg)   \nonumber \\[-8pt]
     &+ \int_0^{\hat{\omega}_0}  \frac{d \hat{\omega}}{(1-\omega)^2}  \int dX \int d y_1 \bigg(I_{\tilde{s}}^s + I_{s'}^s  \nonumber \\[-8pt]
     & \qquad +  X \frac{d}{d s} \Big(\tilde{C}_1^{(4,\ \Phi)}  e^{-s/M^2}  (-1 + 2 \log (\tilde{a}_s X (s - s'_s)))\Big)|_{s = \tilde{s}} \bigg) \frac{\Phi_B^\pm(\omega)}{2} \nonumber \\[-8pt] 
     &  -  \int_0^{s_0} ds\  e^{-s/M^2} \int \frac{d\hat{\omega}}{ (1- \hat{\omega})^2} \int dX \int dy_1 2 X \tilde{C}_1^{(4,\ \Phi)}  \bigg[\frac{1}{(s - \tilde{s})^2}\bigg]_{++} \Theta(s'_s-s) \frac{\Phi_B^\pm(\omega)}{2} \nonumber \\[-8pt] 
    & + \int \frac{d \hat{\omega}}{(1-\omega)} \int dX \int d y_1 (\frac{I_{\tilde{s}}^s}{a_s} + \frac{I_{s'}^s}{a_s}) (\tilde{C}_0^{(4,\ \Phi)} \to (y_1 \tilde{C}_0^{(5,\ \Phi)} + (1-y_1)\tilde{C}_0^{(6,\ \Phi)}) \nonumber \\[-8pt] & \text{and}\ \tilde{s} \leftrightarrow s') \frac{\Phi_B^\pm(\omega)}{2} - \int \frac{d\hat{\omega}}{(1- \hat{\omega})}  \int dX \int dy_1 X\bigg(\tilde{C}_1^{(0,\ \phi)} \phi_B^+(\omega)  + (\tilde{C}_1^{(1,\ \Phi)} \nonumber \\[-8pt] & +\tilde{C}_1^{(2,\ \Phi)} +\tilde{C}_1^{(3,\ \Phi)} - 5(y_1 \tilde{C}_2^{(5,\ \Phi)}  +(1-y_1) \tilde{C}_2^{(6,\ \Phi)})e^{\frac{-\tilde{s}}{M^2}} \frac{\Phi_B^\pm(\omega)}{2} \bigg)
    \Bigg\}\, ,
    \label{eq:factA2f}
\end{align}
where,
\begin{align*}
    I_{\tilde{s}}^s = &  \Theta(\tilde{s} - m_s^2) \Theta(s_0 - \tilde{s}) \frac{d}{ds} \left( e^{-s/M^2} \frac{\tilde{C}_0^{(4,\ \Phi)}}{a_s \ s - S_s} \right) \Bigg|_{s = \tilde{s}(\omega)}\,, \nonumber \\[-8pt]
    I_{s'}^s = &  \frac{e^{-s'_s/M^2}}{|a_s|} \Theta(s'_s - m_s^2) \Theta(s_0 - s'_s)   \frac{\tilde{C}_0^{(4,\ \Phi)}}{(s'_s -  \tilde{s})^2} \,,\nonumber \\[-8pt] 
    s'_s =& \tilde{s} - \frac{1-y_1}{y_1} (1- \hat{\omega})m_B^2 - \frac{X(1-y_1)}{1-X} \frac{q^2}{1-\hat{\omega}}\,, \nonumber \\[-8pt]
    S_s =& a_s s'_s, \ \ a_s = y_1 (X-1)(1-\hat{\omega})\,.
\end{align*}
}
\begin{align}
    I^{(A3)}(M^2, q^2) &= \frac{1}{(4\pi)^2} \frac{1}{ m_B^2} \Bigg\{ \int_0^{\hat{\omega}_0} d\hat{\omega}\, e^{\frac{- \tilde{s}}{M^2}} \,  \int_0^1 dx \ \Bigg[ ( \log((x^2 - x)(1-\hat{\omega})^2 \tilde{m}_B^2) ) \Bigg( \frac{\tilde{A}_0^{(0,\phi)}}{(1- \hat{\omega})^3}  \phi_B^+(\omega) \nonumber \\[-8pt] & +  \  \frac{ 2 \tilde{A}_1^{(0,\phi)} (x^2-x) m_B^2}{(1- \hat{\omega})} \phi_B^+(\omega) + \frac{\tilde{A}_0^{(2,\Phi)} + \tilde{A}_0^{(3,\Phi)} + \tilde{A}_0^{(5,\Phi)}}{(1- \hat{\omega})^3} \frac{\Phi_B^\pm(\omega)}{2} \nonumber \\&  +    \frac{ 2 (\tilde{A}_1^{(2,\Phi)} + \tilde{A}_1^{(5,\Phi)}) (x^2 - x) m_B^2}{(1- \hat{\omega})} \frac{\Phi_B^\pm(\omega)}{2} +  \Big(- \frac{1}{M^2} \frac{\tilde{A}_0^{(4,\Phi)} }{(1- \hat{\omega})^4}  + \frac{\tilde{A}_0^{'\ (4,\Phi)}}{(1 - \hat{\omega})^4}  \Big)  \frac{\Phi_B^\pm(\omega)}{2} \nonumber \\[-8pt] &  +  \Big(- \frac{1}{M^2} \frac{\tilde{A}_1^{(4,\Phi)} }{(1- \hat{\omega})^2}  + \frac{\tilde{A}_1^{'\ (4,\Phi)}}{(1 - \hat{\omega})^2}  \Big) 2 (x^2 - x) m_B^2 \frac{\Phi_B^\pm(\omega)}{2} + 2 x \  \frac{\tilde{A}_1^{(6,\Phi)}}{(1- \hat{\omega})^3} \frac{\Phi_B^\pm(\omega)}{2} \Bigg) \nonumber \\[-8pt] & - \frac{1}{1 - x} \  \frac{\tilde{A}_0^{(6,\Phi)} }{(1- \hat{\omega})^5 m_B^2}  \frac{\Phi_B^\pm(\omega)}{2} - \frac{1}{(1- \hat{\omega})} \bigg( \bigg(\tilde{A}_1^{(0,\phi)}\phi_B^+(\omega) +(\tilde{A}_1^{(2,\Phi)} + \tilde{A}_1^{(5,\Phi)}  + \frac{x \tilde{A}_1^{(6,\Phi)}}{(1-\hat{\omega})^2}) \nonumber \\[-8pt] & \times \frac{\Phi_B^\pm(\omega)}{2} \bigg) (x^2-x) m_B^2 \bigg)   -   \Big(- \frac{1}{M^2} \frac{\tilde{A}_1^{(4,\Phi)} }{(1- \hat{\omega})^2} + \frac{\tilde{A}_1^{'\ (4,\Phi)}}{(1 - \hat{\omega})^2}  \Big) (x^2 - x) m_B^2 \frac{\Phi_B^\pm(\omega)}{2} \Bigg] \Bigg\}\,,
\end{align}

%%%%

\begin{align}
   I^{(B1)}(M^2, q^2) =&\   \frac{-1}{(4\pi)^2} \frac{1}{m_b^2} \int_0^{\hat{\omega}_0} d\hat{\omega}\, e^{\frac{-\tilde{s}}{M^2}}  \frac{1}{(1 - \hat{\omega})} \int_0^1 dx \nonumber \\[-8pt] 
    &\Bigg\{\Bigg[  \big( \bar{A}_0^{(0,\phi)}   + 2\bar{A}_1^{(0,\phi)} m_b^2 \big)  \phi_B^+(\omega)  + \Bigg( \bar{A}_0^{(1,\Phi)}   + 2\bar{A}_1^{(1,\Phi)}   m_b^2  + \bigg( -\frac{\bar{A}_0^{(2,\Phi)}}{(1 - \hat{\omega})M^2}  \nonumber\\[-8pt] 
    &+ \frac{\bar{A}_0^{'\ (2,\Phi)}}{(1 - \hat{\omega})} \bigg) + 2 \bigg(- \frac{\bar{A}_1^{(2,\Phi)}}{(1 - \hat{\omega})M^2} + \frac{\bar{A}_1^{'\ (2,\Phi)}}{(1 - \hat{\omega})} \bigg) m_b^2  \Bigg) \nonumber  \frac{\Phi_B^\pm (\omega)}{2} \Bigg] \log(\tilde{m}_b^2) \nonumber \\[-8pt] & - \bigg(\bar{A}_1^{(0,\Phi)} \phi_B^+ (\omega) + (\bar{A}_1^{(1,\Phi)} - \frac{\bar{A}_1^{(2,\Phi)}}{(1 - \hat{\omega})M^2} + \frac{\bar{A}_1^{'\ (2,\Phi)}}{(1 - \hat{\omega})})\frac{\Phi_B^\pm (\omega)}{2} \bigg)   m_b^2 \Bigg\}\,, 
\end{align}
{\allowdisplaybreaks
\begin{align}
    I^{(B2)}(M^2, q^2) =&\ \frac{1}{(4\pi)^2} \Bigg\{ \int_0^{\hat{\omega}_0}   \frac{d\hat{\omega}}{(1-\hat{\omega})} \int dX\int dy_1 \bigg[\frac{1 }{a_b\ (\tilde{s} - s'_b)}  \bigg( e^{-\tilde{s}/M^2} \bigg(\bar{C}_0^{(0,\ \phi)} \phi_B^+(\omega) \nonumber \\[-8pt] &+ \bar{C}_0^{(1,\ \Phi)} \frac{\Phi_B^\pm (\omega)}{2} \bigg) \Theta(s_0 - s'_b) - e^{-s'_b/M^2} \bigg(\bar{C}_0^{(0,\ \phi)} \phi_B^+(\omega) + \bar{C}_0^{(1,\ \Phi)} \frac{\Phi_B^\pm (\omega)}{2} \bigg) \Theta(s_0 - s'_b) \bigg) \nonumber \\[-8pt] & \quad + X (-1 + 2  \log(\tilde{a}_b X(\tilde{s} - s_b')) )    \bigg(\bar{C}_1^{(0,\ \phi)} \phi_B^+(\omega) + \bar{C}_1^{(1,\ \Phi)} \frac{\Phi_B^\pm (\omega)}{2}\bigg) \bigg] \nonumber\\[-8pt]
    & -  \int_0^{s_0} ds \int   \frac{d\hat{\omega}}{(1-\hat{\omega})} \int dX\int dy_1\ e^{\frac{-s}{M^2}}  2 X    \Theta(s'-s) \left[\frac{1}{s-\tilde{s}}\right]_+ \bigg( \bar{C}_1^{(0,\ \phi)} \phi_B^+(\omega) \nonumber \\[-8pt] & \qquad + \bar{C}_1^{(1,\ \Phi)} \frac{\Phi_B^\pm (\omega)}{2} \bigg)  +  \int_0^{\hat{\omega}_0}  \frac{d \hat{\omega}}{(1-\omega)^2} \int dX \int d y_1 \frac{\Phi_B^\pm}{2} \bigg(I_{\tilde{s}}^b + I_{s'}^b \nonumber \\[-8pt] & \qquad -  X \frac{d}{d s} \left(\bar{C}_1^{(2,\ \Phi)}  e^{-s/M^2}\right)_{s = \tilde{s}} + 2 X \frac{d}{d s} \left(\bar{C}_1^{(2,\ \Phi)}  e^{-s/M^2} \log (\tilde{a}_b X (s - s'))\right)_{s = \tilde{s}} \bigg)   \nonumber \\[-8pt] &
     -  \int_0^{s_0} ds  e^{-s/M^2} \int  \frac{d\hat{\omega}}{ (1- \hat{\omega})^2} \int dX \int dy_1 2 X \bar{C}_1^{(2,\ \Phi)}  \Theta(s'-s) \left[\frac{1}{(s-\tilde{s})^2}\right]_{++} \frac{\Phi_B^\pm}{2} \Bigg\}\, ,
\end{align}
with:
\begin{align}
    I_{\tilde{s}}^b = &  \Theta(\tilde{s} - m_s^2) \Theta(s_0 - \tilde{s}) \frac{d}{ds} \left( e^{-s/M^2} \frac{\bar{C}_0^{(2,\ \phi)}}{a_b\ s - S_b} \right) \Bigg|_{s = \tilde{s}(\omega)}\,, \nonumber \\[-8pt]
    I_{s'}^b = &  \frac{e^{-s'_b/M^2}}{|a_b|} \Theta(s'_b - m_s^2) \Theta(s_0 - s'_b)   \frac{\bar{C}_0^{(2,\ \phi)}}{(s'_b -  \tilde{s})^2}\,, \nonumber \\[-8pt]
    s'_b =& m_B^2 \hat{m}_b \big( \hat{m}_b (1 - y_1 -X) + (1 - y_1) (X - 2) \big) + (1 - y_1) q^2 (\hat{m}_b (X-1) -X (1 - y_1) + 1 )\,, \nonumber \\[-8pt]
    S_b =& a_b s'_b, \ \ a_b = \hat{m}_b (1 - y_1) (X-1)\,.
    \nonumber
\end{align}
}

\begin{align}
     I^{(B3)}(M^2, q^2) =&\  \frac{-1}{(4\pi)^2}\frac{1}{m_B^2} \int_0^{\hat{\omega}_0} d\hat{\omega}\ e^{-\frac{\tilde{s}}{M^2}} \frac{1}{(1 - \hat{\omega})^3} \int dx \nonumber\\[-8pt] & \Bigg[- (\bar{B}_0^{(0,\ \phi) }  + 2 \bar{B}_1^{(0,\ \phi) } x^2 m_b^2 ) \log(x^2 \tilde{m}_b^2) \phi_B^+(\omega) - \Bigg( \bar{B}_0^{(1,\Phi)} + \bar{B}_0^{(2,\Phi)} \nonumber \\[-8pt] 
    & + \bigg(- \frac{\bar{B}_0^{(3,\Phi)}}{(1-\hat{\omega})M^2} + \frac{\bar{B}_0^{'\ (3,\Phi)}}{(1-\hat{\omega})}\bigg) + \frac{\bar{B}_0^{(4,\Phi)}}{(1-\hat{\omega})^2 m_B^2} 
     + \bigg( \bar{B}_1^{(1,\Phi)} + \bar{B}_1^{(2,\Phi)} +\nonumber \\[-8pt] &+ \bigg(-\frac{\bar{B}_1^{(3,\Phi)}}{(1-\hat{\omega})M^2} +\frac{\bar{B}_1^{'\ (3,\Phi)}}{(1-\hat{\omega})} \bigg) + \frac{\bar{B}_1^{(4,\Phi)}}{(1-\hat{\omega})^2 m_B^2}  \bigg) x^2 m_b^2\Bigg) \log(x^2 \tilde{m}_b^2) \frac{\Phi_B^\pm (\omega)}{2} \nonumber \\[-8pt] & +  \bigg( \bar{B}_1^{(0,\ \phi)} \phi_B^+(\omega)  + \big( \bar{B}_1^{(1,\Phi)} + \bar{B}_1^{(2,\Phi)} - \frac{\bar{B}_1^{(3,\Phi)}}{(1-\hat{\omega}) M^2}+\frac{\bar{B}_1^{'\ (3,\Phi)}}{(1-\hat{\omega})} \nonumber \\[-8pt] & + \frac{\bar{B}_1^{(4,\Phi)}}{(1-\hat{\omega})^2 m_B^2}  \big) \frac{\Phi_B^\pm (\omega)}{2}  \bigg) x^2 m_b^2 \Bigg] \,.
     \label{eq:factB3}
\end{align}

\subsection{Hard-scattering contributions}
For the hard-scattering diagrams $(C)$, the explicit form of 
$I^{(C1)}$ is given in the main text. The remaining contributions read:
\begin{align}
     I^{(C2)}(M^2, q^2) = & \frac{-1}{(4\pi)^2} \int_{m_s^2}^{s_0} ds\,e^{-s/M^2}  \,\int_0^\infty d\omega \, \int_0^1 dX \int_0^{1} dy_1\, \int _0^{1} dy_2 \, (1 - y_1) \; 
    \nonumber\\[-8pt] &\quad   \Bigg\{ \Bigg(\frac{-1}{M^2} B_0^{{(0,\phi)}}(\omega, X,y_1,y_2,s,q^2) +B_0'^{{(0,\phi)}}(\omega, X,y_1,y_2,q^2) \nonumber \\[-8pt] & - 2 X B_1^{{(0,\phi)}}(\omega, X,y_1,y_2,s,q^2)\Bigg) \frac{\delta(\mathcal{W}_{C2}\, s - \mathcal{X}_{C2})}{\mathcal{W}_{C2}} 
  \;  \phi_B^+(\omega) \nonumber\\[-8pt]
    &+ \Bigg( -\frac{ \big( B_0^{(1,\Phi)} + B_0^{(2,\Phi)} + B_0^{(3,\Phi)}\big)}{M^2}  +\big( B_0^{'(1,\Phi)} + B_0^{'(2,\Phi)} + B_0^{'(3,\Phi)}\big) \nonumber \\[-8pt] & - 2 X \ \big( B_1^{(1,\Phi)} + B_1^{(2,\Phi)} + B_1^{(3,\Phi)}\big)    + \frac{1}{M^4}  \big(y_1 B_0^{(4,\Phi)} + (1-y_1) B_0^{(5,\Phi)} \nonumber \\[-8pt] &+ (1-y_1) (1-y_2) B_0^{(6,\Phi)}\big)   - \frac{1}{M^2} 2 \big(y_1 B_0^{'(4,\Phi)} + (1-y_1) B_0^{'(5,\Phi)}  + (1-y_1) \nonumber \\[-8pt] & \quad \times (1-y_2) B_0^{'(6,\Phi)}\big) + \big(y_1 B_0^{''(4,\Phi)} + (1 - y_1) B_0^{''(5,\Phi)}  + (1 - y_1) (1 - y_2) B_0^{''(6,\Phi)}\big) \nonumber \\[-8pt] &  -  \frac{2 X \ \big( y_1 B_1^{(4,\Phi)} + (1-y_1) B_1^{(5,\Phi)} + (1-y_1)(1-y_2) B_1^{(6,\Phi)} \big)}{M^2} \nonumber \\[-8pt] & + 2 X \big( y_1 B_1^{'(4,\Phi)} + (1-y_1) B_1^{'(5,\Phi)} + (1-y_1) (1-y_2) B_1^{'(6,\Phi)} \big)  + 6 X^2 \big( y_1 B_2^{(4,\Phi)} \nonumber \\[-8pt] & + (1-y_1) B_2^{(5,\Phi)}  + (1-y_1) (1-y_2) B_2^{(6,\Phi)} \big) \Bigg) \frac{\delta(\mathcal{W}_{C2}\, s - \mathcal{X}_{C2})}{\mathcal{W}_{C2} } \frac{\Phi_B^\pm(\omega)}{2} \Bigg\}\,, 
    \label{eq:fig1atw2f}
\end{align}
where
\begin{align}
    (\mathcal{W}_{C2} s - \mathcal{X}_{C2}) =& - \Big[  y_2 \bar{y}_1 (1 -\omega - X(1 - \omega - y_1)) s \nonumber\\[-8pt] & + \left( \bar{X} \bar{y}_1 \omega (q^2 y_2 - m_B^2(1 + \bar{y}_2) ) + \bar{y}_1 \bar{y}_2 (m_B^2 - m_B^2 X \bar{y_1} + q^2 X \bar{y}_1 y_2) + m_B^2 \omega^2 \bar{X}\right) \Big]\,.\nonumber
\end{align}
\allowdisplaybreaks{
\begin{align}
     I^{(C3)}(M^2, q^2) = & \frac{-1}{(4\pi)^2} \int_{m_s^2}^{s_0} ds\,e^{-s/M^2}  \,\int_0^\infty d\omega \, \int_0^1 dX \int_0^{1} dy_1\, \int _0^{1} dy_2 \, (1 - y_1) \; 
    \nonumber\\[-8pt] &\quad   \Bigg\{ \Bigg(\frac{-1}{M^2} A_0^{{(0,\phi)}}(\omega, X,y_1,y_2,s,q^2) +A_0'^{{(0,\phi)}}(\omega, X,y_1,y_2,q^2) \nonumber \\[-8pt] & - 2 X A_1^{{(0,\phi)}}(\omega, X,y_1,y_2,s,q^2)\Bigg) \frac{\delta(\mathcal{W}_{C3}\, s - \mathcal{X}_{C3})}{\mathcal{W}_{C3}} 
  \;  \phi_B^+(\omega) \nonumber\\[-8pt]
    &+ \Bigg( -\frac{ \big( A_0^{(1,\Phi)} + A_0^{(2,\Phi)} + A_0^{(3,\Phi)}\big)}{M^2}  +\big( A_0^{'(1,\Phi)} + A_0^{'(2,\Phi)} + A_0^{'(3,\Phi)}\big) \nonumber \\[-8pt] & - 2 X \ \big( A_1^{(1,\Phi)} + A_1^{(2,\Phi)} + A_1^{(3,\Phi)}\big)   +  \frac{ \big(y_1 A_0^{(4,\Phi)} + (1-y_1) A_0^{(5,\Phi)} + (1-y_1) (1-y_2) A_0^{(6,\Phi)}\big) }{M^4} \nonumber \\[-8pt]& -\frac{ 2 \big(y_1 A_0^{'(4,\Phi)} + (1-y_1) A_0^{'(5,\Phi)} + (1-y_1)(1-y_2) A_0^{'(6,\Phi)}\big) }{M^2} \nonumber \\[-8pt] & + \big(y_1 A_0^{''(4,\Phi)} + (1-y_1) A_0^{''(5,\Phi)} + (1-y_1) (1-y_2) A_0^{''(6,\Phi)}\big)  \nonumber \\[-8pt] & -  \frac{2 X \ \big( y_1 A_1^{(4,\Phi)} + (1-y_1) A_1^{(5,\Phi)} + (1-y_1)(1-y_2) A_1^{(6,\Phi)} \big)}{M^2} \nonumber \\[-8pt] & + 2 X \big( y_1 A_1^{'(4,\Phi)} + (1-y_1) A_1^{'(5,\Phi)} + (1-y_1) (1-y_2) A_1^{'(6,\Phi)} \big)  + 6 X^2 \big( y_1 A_2^{(4,\Phi)} \nonumber \\[-8pt] & + (1-y_1) A_2^{(5,\Phi)}  + (1-y_1) (1-y_2) A_2^{(6,\Phi)} \big) \Bigg) \frac{\delta(\mathcal{W}_{C3}\, s - \mathcal{X}_{C3})}{\mathcal{W}_{C3} } \frac{\Phi_B^\pm(\omega)}{2} \Bigg\}\,, 
    \label{eq:figC3}
\end{align}
}
where
\begin{align}
   \mathcal{W}_{C3}\, s - \mathcal{X}_{C3} = &  m_B^2 \left[-\left((1 - X\,\bar{y_1})\,\bar{y}_1\,\bar{y}_2\right) - \bar{X}\, \bar{y}_1(-2 + y_2)\,\omega - \bar{X}\,\omega^2\right] \nonumber\\[-8pt] &\quad
- \bar{y}_1 y_2 \left[q^2 + s\,X\, \bar{y_1}\,\bar{y_2} - q^2\,\omega + s\bar{X}\,\omega + q^2 X(-\bar{y_1} + \omega)\right]\,.\nonumber 
\end{align}
\allowdisplaybreaks{
\begin{align}
    I^{(C4)}(M^2, q^2) = & \int_{m_s^2}^{s_0} d s\, e^{-s/M^2}\int_{0}^{\infty} d\omega\, \int_0^1 dX \, dy_1 \,  \frac{1}{\hat{\omega}(s - a_{C4}(q^2, \omega))} 
      \nonumber\\[-8pt] & \Bigg\{\Bigg(\frac{C_{0}^{{(0,\phi)}}}{\mathcal{W}_{C4}} \delta(\mathcal{W}_{C4} s - \mathcal{X}_{C4}) -  2 X C_{1}^{{(0,\phi)}}\ \Theta(\mathcal{X}_{C4} - \mathcal{W}_{C4} s)\,  \Bigg)\phi_B^+(\omega) \nonumber \\[-8pt]
      & + \Bigg(\bigg( \big(C_0^{(1,\Phi)} + C_0^{(2,\Phi)} + C_0^{(3,\Phi)}\big)+ \frac{  C_0^{(4,\Phi)} }{\hat{\omega} (s - a_{C4}(q^2,\omega))} + \frac{ \big(y_1 C_0^{(5,\Phi)} + (1-y_1) C_0^{(6,\Phi)} \big) }{M^2}   \nonumber\\[-8pt] & - \big(y_1 C_0^{'(5,\Phi)}+ (1 - y_1) C_0^{'(6,\Phi)} \big)  + 2X \big( y_1 C_1^{(5,\Phi)} + (1-y_1) C_1^{(6,\Phi)} \big)\bigg) \frac{\delta(\mathcal{W}_{C4} s - \mathcal{X}_{C4})}{\mathcal{W}_{C4}} \nonumber\\[-8pt] & - 2 X \bigg( C_1^{(1,\Phi)} + C_1^{(2,\Phi)}  + C_1^{(3,\Phi)} + \frac{ C_1^{(4,\Phi)} }{\hat{\omega} (s - a(q^2,\omega))}+ 3 X \big( y_1 C_2^{(5,\Phi)}  \nonumber \\[-8pt] &\qquad + (1 - y_1) C_2^{(6,\Phi)} \big)\bigg)  \Theta(\mathcal{X}_{C4} - \mathcal{W}_{C4} s)     \Bigg)\frac{\Phi_B^\pm(\omega)}{2} \Bigg\}\,,
\end{align}
}
where
\begin{align}
   \mathcal{W}_{C4}\, s - \mathcal{X}_{C4}  &= -  y_1(\omega \bar{X} + X \bar{y}_1) s-  \bar{X} \bar{\omega} (q^2 y_1 - m_B^2 (\omega - \bar{y}_1))\,.\nonumber
\end{align}

\subsection{Soft-gluon contribution}
For the soft-gluon diagrams $(D)$, the explicit form of 
$I^{(D1)}$ is given in the main text. The remaining contributions read:
\begin{align}
    I^{(D2)} (p^2, q^2) = & \sum_{n=1,2} \int_0^\infty d\omega \int_0^\infty d\xi \frac{1}{[(\omega v -p)^2]^n} \Big[ \tilde{F}_n^{(\Psi_{AV})} (q^2,\omega)(\Psi_A(\omega, \xi ) - \Psi_V(\omega, \xi )) \nonumber \\& + \tilde{F}_n^{(\Psi_{V})}(q^2,\omega) \Psi_V(\omega, \xi ) + \tilde{F}_n^{(X_{A})}(q^2,\omega) \overline{X}_A(\omega, \xi ) + \tilde{F}_n^{(W Y_{A})}(q^2,\omega) \big(\overline{W} (\omega,\xi)\nonumber\\ & + \overline{Y}_A(\omega, \xi)\big)  + \tilde{F}_n^{(\tilde{X}_{A})}(q^2,\omega) \overline{\tilde{X}}_A(\omega, \xi)+ \tilde{F}_n^{(\tilde{Y}_{A})}(q^2,\omega) \overline{\tilde{Y}}_A(\omega, \xi)    \Big]\,.
\end{align}
These contributions are expressed in terms of three-particle $B$-meson DAs and the coefficient functions defined in Appendix~\ref{app:DAs} and~\ref{app:coeff}.

\noindent

\section{Coefficients  in the LCSR integrals}
\label{app:coeff}
This appendix collects the explicit expressions of the coefficient functions $\tilde{B}_i^{(j,\phi/\Phi)}$, $D_i^{(j,\phi/\Phi)}$, and $F_n^{(X)}$ that enter the invariant amplitudes $I^{(A1)}$, $I^{(C1)}$, and $I^{(D1)}$, respectively. The full set of coefficients used in the diagrams of Fig.~\ref{fig:O8gdiag} are given in the ancillary file accompanying this paper.
All coefficients depend on the external and kinematic variables as $$\tilde{X}_i^{(j,\phi/\Phi)} = 
\tilde{X}_i^{(j,\phi/\Phi)}(s,q^2,m_B,m_b,\hat{\omega};\,x~\text{or}~x_2,x_3),\;\;
\hat{m}_b = m_b/m_B,\;\; \hat{\omega} = \omega/m_B\,.$$
The superscript $\phi(\Phi)$ denotes the terms multiplying the twist–$2(3)$ $B$-meson distribution amplitudes defined in Appendix~\ref{app:DAs}. In the analytical expressions presented in this appendix, the strange-quark mass $m_s$ is omitted for brevity; however, it is considered in our numerical evaluations.
\subsection{Coefficients \texorpdfstring{$\tilde{B}_i^{(j,\phi/\Phi)}$}{Btilde}}\label{app:coeff_IA1}
The coefficients used in the
$I^{(A1)}$ function ($\tilde{B}$) are given below: 
{
\allowdisplaybreaks
\begin{align}
\tilde{B}_0^{(0,\phi)} = &4\, m_B\,(2\hat{m}_b-1)\,(x-1)\,x\,(\hat{\omega}-1)\big[ (\hat{\omega}-1)(2\hat{\omega}-5)\,s
 + \big((3-2\hat{\omega})\hat{\omega}+2\big)\,q^{2}
 \nonumber \\[-8pt] & + m_B^{2}\,(\hat{\omega}^{2}+\hat{\omega}-2)\big]\nonumber \, ,\\
\tilde{B}_1^{(0,\phi)} = &12\, m_B\,(2\hat{m}_b-1)\,(\hat{\omega}-1) \nonumber \, ,\\ \tilde{B}_0^{(1,\Phi)} = &-8\,(x-1)\,x\big[ (\hat{\omega}-1)\,s\big(2\hat{m}_b(\hat{\omega}-4)-2\hat{\omega}+5\big)
  + q^{2}\big(-2\hat{m}_b(\hat{\omega}-2)\hat{\omega}+4\hat{m}_b  -3\big) \nonumber \\[-8pt] &
 + m_B^{2}(\hat{\omega}-1)\big(4\hat{m}_b\hat{\omega}+2\hat{m}_b-\hat{\omega}-2\big)\big]\, , \nonumber \\ \tilde{B}_1^{(1,\Phi)} = &-24\,(2\hat{m}_b-1)\nonumber \, , \\ 
\tilde{B}_0^{(2,\Phi)} = & -8\,x\big[-(2\hat{m}_b+1)(\hat{\omega}-1)\,s
 + q^{2}\big(2\hat{m}_b(\hat{\omega}-2)-3\hat{\omega}+4\big)
 + m_B^{2}(\hat{\omega}-1) \nonumber\\[-8pt] & \big(6\hat{m}_b\hat{\omega}-4\hat{m}_b-3\hat{\omega}+4\big)\big] \nonumber \, ,\\
\tilde{B}_1^{(2,\Phi)} = &\ 0 \nonumber \, ,\\
\tilde{B}_0^{(3,\Phi)} = &-8\,(x-1)\,x\Big\{ -(\hat{\omega}-1)\,q^{2}\big[ s\big(2\hat{m}_b(\hat{\omega}-4)-5\hat{\omega}+11\big)
 + m_B^{2}\big(2\hat{m}_b\big(2(\hat{\omega}-2)\hat{\omega}^{2}+\hat{\omega}+4\big)
 \nonumber \\[-8pt] & + \hat{\omega}\big((7-2\hat{\omega})\hat{\omega}-3\big)-8\big)\big] 
+ (\hat{\omega}-1)^{2}\,s\big[(5-2\hat{\omega})\,s
 + 2 m_B^{2}\big(\hat{m}_b(\hat{\omega}-1)(2\hat{\omega}-5)
 \nonumber \\[-8pt] & - (\hat{\omega}-4)\hat{\omega}-6\big)\big] 
+ q^{4}\left(2\hat{m}_b\big((\hat{\omega}-2)\hat{\omega}-2\big)
 + \hat{\omega}\big(2(\hat{\omega}-3)\hat{\omega}+3\big)+4\right)
 \nonumber \\[-8pt] & + m_B^{4}(\hat{\omega}-1)^{2}(\hat{\omega}+2)\big(2\hat{m}_b(\hat{\omega}-1)-\hat{\omega}+2\big)
\Big\} \nonumber \, , \\
\tilde{B}_1^{(3,\Phi)} = &-24\big[-q^{2}(2\hat{m}_b+\hat{\omega}-2)  - (\hat{\omega}-1)\,s 
 + m_B^{2}(\hat{\omega}-1)\big(2\hat{m}_b(\hat{\omega}-1)-\hat{\omega}+2\big)\big] \nonumber \, , \\
 \tilde{B}_0^{(4,\Phi)} = & 8\,(x-1)^{2}x\Big\{ -(\hat{\omega}-2)(\hat{\omega}-1)\,q^{2}\big[ s\big(-2\hat{m}_b(\hat{\omega}-4) +5\hat{\omega}-11\big)  \nonumber \\[-8pt] &
 + m_B^{2}\left(2\hat{m}_b\big((\hat{\omega}-1)\hat{\omega}-3\big)
 + (3-4\hat{\omega})\hat{\omega}+7\right)\big] 
+ (\hat{\omega}-1)^{2}\,s\big[(\hat{\omega}-2)(2\hat{\omega}-5)\,s \nonumber \\[-8pt] &
 + m_B^{2}\left(2\hat{m}_b(\hat{\omega}-1)(2\hat{\omega}-5)
 + (16-3\hat{\omega})\hat{\omega}-19\right)\big] - (\hat{\omega}-2)\,q^{4}\big(2\hat{m}_b\big((\hat{\omega}-2)\hat{\omega}-2\big)
 \nonumber \\[-8pt] & + \hat{\omega}\big(2(\hat{\omega}-3)\hat{\omega}+3\big)+4\big)
 + m_B^{4}(\hat{\omega}-1)^{2}(\hat{\omega}+2)\big(2\hat{m}_b(\hat{\omega}-1)-2\hat{\omega}+3\big)
\Big\} \nonumber \, , \\
\tilde{B}_1^{(4,\Phi)} = & 4\Big\{
 -(\hat{\omega}-1)\,s\big(\hat{\omega}\,(3\hat{m}_b(x-1)-5x+7)+4\hat{m}_b+14x-15\big) \nonumber \\[-8pt]
&
 + q^{2}\big(\hat{\omega}^{2}(3\hat{m}_b(x-1)+5x-3)
 + \hat{\omega}\,(\hat{m}_b(11x-7)-21x+18) - 26\hat{m}_b x  + 24\hat{m}_b \nonumber \\[-8pt] & + 25x - 23\big) 
 + m_B^{2}(\hat{\omega}-1)\big(\hat{m}_b\,(3x(7\hat{\omega}-6)-15\hat{\omega}+16)
 - 14x\hat{\omega}+23x+13\hat{\omega}-21\big) \nonumber\\[-8pt]
& - 2\big(\hat{m}_b\big(\hat{\omega}\,(2(x-1)\hat{\omega}+x+3)-2\big)
 + x\hat{\omega}(4\hat{\omega}-5)+x-\hat{\omega}+1\big)
\Big \} \nonumber \, .
\end{align}
}

\subsection{Coefficients \texorpdfstring{$D_i^{(j,\phi/\Phi)}$}{D}}\label{app:coeff_IC1}
The coefficients used in the $I^{(C1)}$ function ($D$) are given below: 
{\allowdisplaybreaks
\begin{align}
    D_0^{{(0,\phi)}} (s,q^2) = & -4 \, m_B\Big(\hat{\omega}\Big( x_3 q^{2}\big[\hat{\omega} (x_2+x_3-1)\big(6 \hat{m}_b (x_2+x_3-1)-3 x_2-3 x_3+4\big) \nonumber \\[-8pt]&-(x_3-1)\big((6 \hat{m}_b-3) x_2+6 \hat{m}_b x_3-3 x_3+1\big)\big] +\, m_B^{2}\big[3 (2 \hat{m}_b-1)\,\hat{\omega}^{2} \nonumber \\[-8pt] & \qquad (x_2+x_3-1)^{2} (x_2+x_3) -\,\hat{\omega} (x_2+x_3-1)\big(3 (2 \hat{m}_b-1) x_2 (2 x_3-1) \nonumber \\[-8pt] & +2 x_3 (6 \hat{m}_b x_3-3 \hat{m}_b-3 x_3+2)\big) +\, (x_3-1) x_3 \big((6 \hat{m}_b-3) x_2+6 \hat{m}_b x_3 \nonumber \\[-8pt] & -3 x_3+1\big)\big]\Big) -\, x_3\, s\,\big(\hat{\omega} (x_2+x_3-1)-x_3+1\big)\big(\hat{\omega} (6 \hat{m}_b (x_2-1)-3 x_2 \nonumber \\[-8pt] & +2)+3 (2 \hat{m}_b-1) x_3 (\hat{\omega}-1)\big)\Big) \nonumber \, , \\
    D_1^{{(0,\phi)}} (s,q^2) = & -6 \, m_B (2 \hat{m}_b-1)\,\big((3 x_2-4)\,\hat{\omega}+3 x_3(\hat{\omega}-1)+2\big) \nonumber \, , \\
    D_2^{{(0,\phi)}} (s,q^2) = & 0 \nonumber \, , \\
    D_0^{{(1,\phi)}} (s,q^2) = & 8 \Big(-(4 \hat{m}_b-1)\, x_3\, s\,\big(\hat{\omega} (x_2+x_3-1)-x_3+1\big) +\, x_3 q^{2}\big((4 \hat{m}_b-3)\,\hat{\omega}  (x_2+x_3-1)\nonumber \\[-8pt] &-2 (\hat{m}_b-1)(x_3-1)\big) +\, m_B^{2}\big(\hat{\omega} (x_2+x_3-1)-x_3+1\big)\big(3(2 \hat{m}_b-1)\,\hat{\omega} (x_2+x_3)\nonumber \\[-8pt] & -2 (\hat{m}_b-1) x_3\big)\Big) \nonumber \, , \\
    D_1^{{(1,\phi)}} (s,q^2) = & 12 \,(2 \hat{m}_b-1) \nonumber \, , \\
    D_2^{{(1,\phi)}} (s,q^2) = & 0 \nonumber \, , \\
    D_0^{{(2,\phi)}} (s,q^2) = & 24 \Big( x_3 s\big(\hat{\omega} (-2 \hat{m}_b x_2+2 \hat{m}_b+x_2)+x_3(-2 \hat{m}_b \hat{\omega}+2 \hat{m}_b+\hat{\omega}-1)\big) +\, x_3 \hat{\omega} q^{2}\big(2 \hat{m}_b (x_2 \nonumber \\[-8pt] & +x_3-1)-x_2-x_3+2\big) +\, m_B^{2}\hat{\omega}\big((2 \hat{m}_b-1)\,\hat{\omega} (x_2+x_3-1)(x_2+x_3) \nonumber \\[-8pt] & +x_3(-2 \hat{m}_b x_2-2 \hat{m}_b x_3+x_2+x_3-1)\big)\Big) \nonumber \, , \\
    D_1^{{(2,\phi)}} (s,q^2) = & 24 \,(2 \hat{m}_b-1) \nonumber \, , \\
    D_2^{{(2,\phi)}} (s,q^2) = & 0 \nonumber \, , \\
    D_0^{{(3,\phi)}} (s,q^2) = & -8 \Big( x_3 s\Big[ x_3 q^{2}\Big(\hat{\omega}\big(2 \hat{m}_b (5 x_2 x_3-3 x_2+5 x_3^{2}-7 x_3+2)-6 x_3 (x_2+x_3)+3 x_2+10 x_3-4\big) \nonumber \\[-8pt]& - \,4 (2 \hat{m}_b-1)\,\hat{\omega}^{2} (x_2+x_3-1)^{2}-2 (\hat{m}_b-1) (x_3-1) x_3\Big)-\, 2 m_B^{2}\big(\hat{\omega} (x_2+x_3-1)-x_3 \nonumber \\[-8pt] & +1\big)\Big( \hat{\omega}^{2} (x_2+x_3-1)\big(\hat{m}_b (5 x_2+5 x_3-3)-2 x_2-2 x_3+1\big) +\, x_3 \hat{\omega}\big(\hat{m}_b(-6 x_2-6 x_3+4) \nonumber \\[-8pt] & +3 (x_2+x_3-1)\big)+(\hat{m}_b-1) x_3^{2}\Big)\Big] +\, \hat{\omega}\Big[ x_3 q^{2}\Big( x_3 q^{2}\big(\hat{\omega} (x_2+x_3-1)\big(4 \hat{m}_b (x_2+x_3-1)\nonumber \\[-8pt] & -3 x_2-3 x_3+4\big)-2 (\hat{m}_b-1) (x_3-1)(x_2+x_3)\big) +\, m_B^{2}\Big( 2 \hat{\omega}^{2} (x_2+x_3-1)^{2}\big(\hat{m}_b(5 x_2 \nonumber \\[-8pt] & +5 x_3-3)-3 x_2-3 x_3+2\big)-\, \hat{\omega} (x_2+x_3-1)\Big(x_2\big(2(7 \hat{m}_b-5) x_3-8 \hat{m}_b+5\big) \nonumber \\[-8pt] & +2 x_3\big(7 \hat{m}_b x_3-6 \hat{m}_b-5 x_3+5\big)-1\Big) +\, 4 (\hat{m}_b-1) (x_3-1) x_3 (x_2+x_3)\Big)\Big) \nonumber \\[-8pt] & +\, m_B^{4}\big(\hat{\omega} (x_2+x_3-1)-x_3+1\big)\Big(2 (\hat{m}_b-1) x_3^{2} (x_2+x_3)+3 (2 \hat{m}_b-1)\,\hat{\omega}^{2} (x_2+x_3 -1)^{2} \nonumber \\[-8pt] & (x_2+x_3) -\, x_3 \hat{\omega} (x_2+x_3-1)\big((8 \hat{m}_b-5) x_2+8 \hat{m}_b x_3-5 x_3+1\big)\Big)\Big] +\, x_3^{2} s^{2}\, \nonumber \\[-8pt] & \qquad \big(\hat{\omega} (x_2+x_3-1)-x_3+1\big)\big(4 \hat{m}_b (x_2-1) \hat{\omega}+(4 \hat{m}_b-1) x_3 (\hat{\omega}-1)-x_2 \hat{\omega}\big)\Big)
 \nonumber \, , \\
    D_1^{{(3,\phi)}} (s,q^2) = & -4 \Big(- s\Big( x_3\big(\hat{m}_b (16 x_2 \hat{\omega}-15 \hat{\omega}+7)+\hat{\omega}\big(2 x_2 (\hat{\omega}-3)-3 \hat{\omega}+3\big)-2\big) \nonumber \\[-8pt] & +\, (x_2-2)\,\hat{\omega}\big(\hat{m}_b+(x_2-1)\hat{\omega}+1\big)+x_3^{2}(\hat{\omega}-1)(16 \hat{m}_b+\hat{\omega}-5)\Big) \nonumber \\[-8pt] &+\, q^{2}\Big(\hat{\omega}\big(\hat{m}_b(16 x_2 x_3+x_2+x_3(16 x_3-15)-2)-12 x_3 (x_2+x_3)+x_2+15 x_3+2\big) \nonumber \\[-8pt] & -\, 3 (\hat{m}_b-1) x_3 (2 x_3-1)+\hat{\omega}^{2} (x_2+x_3-2)(x_2+x_3-1)\Big) +\, 2 \hat{\omega}^{2} (x_2+x_3-1) \nonumber \\[-8pt] & \qquad \Big(m_B^{2}\big(\hat{m}_b(11 x_2+11 x_3-8)-5 x_2-5 x_3+3\big)+2\big(8 \hat{m}_b (x_2+x_3-1)-5 x_2 \nonumber \\[-8pt] & -5 x_3+6\big)\Big) +\, \hat{\omega}\Big(-\hat{m}_b\big(m_B^{2}(x_2 (28 x_3-11)+x_3 (28 x_3-33)+6)+40 x_3 (x_2+x_3) \nonumber \\[-8pt] & -4(7 x_2+15 x_3-6)\big) +\, m_B^{2}\big(4 x_2(4 x_3-1)+x_3 (16 x_3-21)\big)+26 x_3 (x_2+x_3) \nonumber \\[-8pt] & -20 x_2-42 x_3+20\Big) +\, x_3\big(3 m_B^{2} (\hat{m}_b-1) (2 x_3-1)+2 \hat{m}_b (6 x_3-7)-8 x_3+10\big)\Big) \nonumber \, , \\
    D_2^{{(3,\phi)}} (s,q^2) = & -12 \,(2 \hat{m}_b-1) \nonumber \, , \\
    D_0^{{(4,\phi)}} (s,q^2) = & 8 \Big( s\Big((x_3-1) x_3 q^{2}\big(4 (2 \hat{m}_b-1)\,\hat{\omega}^{2} (x_2+x_3-1)^{2} +\, \hat{\omega}\big(2 \hat{m}_b\,(x_2(3-5 x_3) \nonumber \\[-8pt] & +(7-5 x_3) x_3-2)+6 x_3 (x_2+x_3)-3 x_2-10 x_3+4\big) +\, 2 (\hat{m}_b-1) (x_3-1) x_3\big) \nonumber \\[-8pt] & +\, m_B^{2}\big(\hat{\omega} (x_2+x_3-1)-x_3+1\big)\Big[ x_3 \hat{\omega}\big(3(2 x_3-1)(x_2+x_3-1) -\,2 \hat{m}_b(6 x_2 x_3 \nonumber \\[-8pt] & -3 x_2+6 x_3^{2}-7 x_3+1)\big) +\, \hat{\omega}^{2} (x_2+x_3-1)\big(x_2(10 \hat{m}_b x_3-4 \hat{m}_b-4 x_3+1) \nonumber \\[-8pt] & +x_3(10 \hat{m}_b (x_3-1)-4 x_3+3)\big) +\, 2 (\hat{m}_b-1) (x_3-1) x_3^{2}\Big]\Big) +\, \hat{\omega}\Big( m_B^{2} q^{2}\Big[-\hat{\omega}^{2} \nonumber \\[-8pt] & \qquad (x_2+x_3-1)^{2}\big(x_2(10 \hat{m}_b x_3-4 \hat{m}_b-6 x_3+3)+x_3(10 \hat{m}_b (x_3-1)-6 x_3+7)\big) \nonumber \\[-8pt] & +\, (x_3-1)\,\hat{\omega} (x_2+x_3-1)\Big(2 x_2(7 \hat{m}_b x_3-\hat{m}_b-5 x_3+1) +\, x_3\big(2(7 \hat{m}_b-5) x_3-6 \hat{m}_b+7\big)\Big) \nonumber \\[-8pt] & -\, 4 (\hat{m}_b-1) x_3 (x_3-1)^{2} (x_2+x_3)\Big] -\, (x_3-1) x_3 q^{4}\big(\hat{\omega} (x_2+x_3-1)(4 \hat{m}_b (x_2+x_3-1) \nonumber \\[-8pt] & -3 x_2-3 x_3+4)-2 (\hat{m}_b-1) (x_3-1) (x_2+x_3)\big) -\, m_B^{4}\big(\hat{\omega} (x_2+x_3-1) \nonumber \\[-8pt] & -x_3+1\big)\Big(3(2 \hat{m}_b-1)\,\hat{\omega}^{2} (x_2+x_3-1)^{2} (x_2+x_3) -\, \hat{\omega} (x_2+x_3-1)\big(x_2(8 \hat{m}_b x_3 \nonumber \\[-8pt] & -2 \hat{m}_b-5 x_3+2)+x_3(8 \hat{m}_b x_3-2 \hat{m}_b-5 x_3+3)\big) +\, 2 (\hat{m}_b-1) (x_3-1) x_3 (x_2+x_3)\Big)\Big) \nonumber \\[-8pt] & -\, (x_3-1) x_3\, s^{2}\,\big(\hat{\omega} (x_2+x_3-1)-x_3+1\big)\big(4 \hat{m}_b (x_2-1) \hat{\omega}+(4 \hat{m}_b-1) x_3 (\hat{\omega}-1)-x_2 \hat{\omega}\big)\Big) \nonumber \, , \\
    D_1^{{(4,\phi)}} (s,q^2) = & -4 \Big(- s\Big(\hat{m}_b \hat{\omega}\big(2 x_2 (8 x_3-5)+2 x_3 (8 x_3-13)+11\big)-2 \hat{m}_b\big(x_3(8 x_3-9)+4\big) \nonumber \\[-8pt] & +\, \hat{\omega}^{2} (x_2+x_3-2)(x_2+x_3-1)+x_3 \hat{\omega} (-6 x_2-6 x_3+5)+3 (x_2-1)\hat{\omega} \nonumber \\[-8pt] & + x_3(5 x_3-4)+2\Big) +\, q^{2}\Big(\hat{\omega}\big(\hat{m}_b\big(2 x_2(8 x_3-5)+2 x_3(8 x_3-13)+11\big) \nonumber \\[-8pt] & -12 x_3 (x_2+x_3)+9 x_2+23 x_3-9\big) -\, 2 (\hat{m}_b-1) (x_3-1) (3 x_3-2)+\hat{\omega}^{2} (x_2 \nonumber \\
&\qquad +x_3-2)(x_2+x_3-1)\Big) +\, 2 \hat{\omega}^{2} (x_2+x_3-1)\Big(m_B^{2}\big(\hat{m}_b (11 x_2+11 x_3-8)  \nonumber \\[-8pt]
& -5 x_2-5 x_3+3\big)+2\big(8 \hat{m}_b (x_2+x_3-1)-5 x_2-5 x_3+6\big)\Big) +\, \hat{\omega}\Big(m_B^{2}\big(\hat{m}_b\big(x_2(18  \nonumber \\[-8pt]
& \qquad -28 x_3)+4(10-7 x_3) x_3-17\big)+16 x_3 (x_2+x_3)-11 x_2-28 x_3+11\big)  \nonumber \\[-8pt]
& +\, 2\big(-2 \hat{m}_b\big(2 x_2 (5 x_3-3)+2 x_3(5 x_3-7)+5\big)+13 x_3 (x_2+x_3)-8 x_2-19 x_3+8\big)\Big)  \nonumber \\[-8pt]
& +\, 2 (x_3-1)\big(m_B^{2}(\hat{m}_b-1)(3 x_3-2)+6 \hat{m}_b x_3-4 x_3\big)\Big) \nonumber \, , \\
    D_2^{{(4,\phi)}} (s,q^2) = &  -12 \,(2 \hat{m}_b-1) \nonumber \, .  
\end{align}
}
\subsection{Coefficients \texorpdfstring{${F}_n^{(X)}$}{F}}\label{app:coeff_ID1}
The coefficients used in the $I^{(D1)}$ function are given below: 
\allowdisplaybreaks{
\begin{align}
    & F_1^{(\Psi_{AV})} (q^2,\omega) =  \frac{3}{m_b^2} (1 - \hat{\omega})m_B \left( 2 \hat{m}_b - 1 \right)\, , \;\;\quad F_2^{(\Psi_{VA})} (q^2,\omega)=0 \nonumber \, , \\
   & F_1^{(\Psi_{V})} (q^2,\omega) = \frac{6}{m_b^2} (1 - \hat{\omega}) m_B \left(2 \hat{m}_b - 1 \right)\, , \;\;\quad F_2^{(\Psi_{V})} (q^2,\omega) = 0 \nonumber \, ,  \\
   & F_1^{(X_{A})} (q^2,\omega) = - \frac{2}{m_b^2} (1-2 \hat{m}_b)\, ,\;\;\quad  \ F_2^{(X_{A})} (q^2,\omega) = - \frac{1}{m_b^2} (1-2 \hat{m}_b) ((1 - \hat{\omega})^2 m_B^2 - q^2) \nonumber \, ,  \\
   & F_1^{(W\,Y_{A})} (q^2,\omega) = \frac{-4}{(1 - \hat{\omega})^2 m_B^2} \, ,  \;\;\quad  
   F_2^{(W\,Y_{A})} (q^2,\omega) = \frac{4 \left( (1 - \hat{\omega})^3 m_B^2+ q^2(2 \hat{\omega} - 1)\right)}{(1 - \hat{\omega})^2 m_B^2}  \nonumber\,,\\
   & F_1^{(\tilde{X}_{A})} (q^2,\omega) =  F_1^{(X_{A})} (q^2,\omega) \, ,  \;\;\quad
   F_2^{(\tilde{X}_{A})} (q^2,\omega) = F_2^{(X_{A})} (q^2,\omega) \nonumber \, ,  \\
   & F_1^{(\tilde{Y}_{A})} (q^2,\omega) = F_1^{(WY_{A})} (q^2,\omega) \, ,  \;\;\quad               F_2^{(\tilde{Y}_{A})} (q^2,\omega) = F_2^{(WY_{A})} (q^2,\omega)\, .  \nonumber
\end{align}
}
We compared our results for these diagrams with the ones obtained earlier in~\cite{Khodjamirian:2012rm} and found a full agreement between expressions
(up to certain twist-3 DAs that were not included yet in that computation).
The only difference is  in the coefficient  $F_1^{(X_{A})} (q^2,\omega)$, which, 
according to \cite{Khodjamirian:2012rm}
is equal to $\frac{-1}{m_b^2}(2 + \hat{m}_b)$.

\bibliography{references}

\end{document}